\documentclass{aa}
\usepackage{txfonts}
\usepackage{graphics}
\usepackage{epsfig}
\begin{document}

\title{The VMC Survey - I. Strategy and First Data}

\author{M.-R.~L. Cioni\inst{1, 2,}\thanks{Research Fellow of the Alexander von Humboldt Foundation}
	\and G.~Clementini\inst{3}
	\and L.~Girardi\inst{4}
	\and R.~Guandalini\inst{1}	
	\and M.~Gullieuszik\inst{5}
	\and B.~Miszalski\inst{1}
	\and M.-I.~Moretti\inst{6}
	\and V.~Ripepi\inst{7}
	\and S.~Rubele\inst{4}
                \and G.~Bagheri\inst{1}
	\and K.~Bekki\inst{8}
                \and  N.~Cross\inst{9}
	\and W.~J.~G.~de~Blok\inst{10}
	\and R.~de~Grijs\inst{11}
	\and J.~P.~Emerson\inst{12}
	\and C.~J.~Evans\inst{13}
	\and B.~Gibson\inst{14}
	\and E.~Gonzales-Solares\inst{15}
	\and M.~A.~T.~Groenewegen\inst{5}
	\and M. Irwin\inst{15}
	\and V.~D.~Ivanov\inst{16}
                \and J.~Lewis\inst{15}
	\and M.~Marconi\inst{7}
                \and J.-B.~Marquette\inst{17, 18}
	\and C.~Mastropietro\inst{19}
	\and B.~Moore\inst{20}	
	\and R.~Napiwotzki\inst{1}		
	\and T.~Naylor\inst{21}
	\and J.~M.~Oliveira\inst{22}
                \and M.~Read\inst{9}
                \and E.~Sutorius\inst{9}
	\and J.~Th.~van Loon\inst{22}
	\and M.~I.~Wilkinson\inst{23}
	\and P.~R.~Wood\inst{24}
}

\offprints{m.cioni@herts.ac.uk}

\institute{
	University of Hertfordshire, Physics Astronomy and Mathematics, Hatfield AL10 9AB, United Kingdom
	\and University Observatory Munich, Scheinerstrasse 1, 81679 M\"{u}nchen, Germany
	\and INAF, Osservatorio Astronomico di Bologna, Via Ranzani 1, 40127 Bologna, Italy
	\and INAF, Osservatorio Astronomico di Padova, Vicolo dell'Osservatorio 5, 35122 Padova, Italy
	\and Royal Observatory of Belgium, Ringlaan 3, 1180 Ukkel, Belgium
	\and University of Bologna, Department of Astronomy, Via Ranzani 1, 40127 Bologna, Italy
	\and INAF, Osservatorio Astronomico di Capodimonte, via Moiariello 16, 80131 Napoli, Italy
	\and ICRAR, M468, University of Western Australia, 35 Stirling Hwy, Crawley 6009, Western Australia
	\and University of Edinburgh, Institute for Astronomy, Blackford Hill, Edinburgh EH9 3HJ, United Kingdom
	\and University of Cape Town, Private Bag X3, Rondebosch 7701, South Africa
	\and Peking University, Kavli Institute for Astronomy and Astrophysics, Yi He Yuan Lu 5, Hai Dian District, Beijing 100871, China
	\and Queen Mary University of London, Mile End Road, London E1 4NS, United Kingdom
	\and UK Astronomy Technology Centre, Blackford Hill, Edinburgh EH9 3HJ, United Kingdom
	\and Centre for Astrophysics, University of Central Lancshire, Preston PR1 2HE, United Kingdom
	\and University of Cambridge, Institute of Astronomy, Madingley Rd, Cambridge CB3 0HA, United Kingdom
	\and European Southern Observatory, Av. Alonso de C\'{o}rdoba 3107, Casilla 19, Santiago, Chile
         \and UPMC Univ. Paris 06, UMR7095, Institut d'Astrophysique de Paris, 75014 Paris, France
         \and CNRS, UMR7095, Institut d'Astrophysique de Paris, 75014 Paris, France
	\and LERMA, Observatoire de Paris, UPMC, CNRS, 61 Av. de l'Observatoire, 75014 Paris, France
	\and University of Zurich, Institute for Theoretical Physics, 8057 Zurich, Switzerland
	\and University of Exeter, School of Physics, Stocker Road, Exeter EX4 4QL, United Kingdom
	\and University of Keele, School of Physical and Geographical Sciences, Staffordshire ST5 5BG, United Kingdom
	\and University of Leicester, University Road, Leicester LE1 7RH, United Kingdom	
	\and Mount Stromlo Observatory, RSAA, Cotter Road, Weston Creek, ACT 2611, Australia	
}

\date{Received 12 November 2010 / Accepted 16 December 2010}

\titlerunning{The VMC Survey - I.}

\authorrunning{Cioni et al.}

\abstract{The new VISual and Infrared Telescope for Astronomy (VISTA) has started operations. Over its first five years it will 
be collecting data for six Public Surveys, one of which is the
near-infrared $YJK_\mathrm{s}$ VISTA survey of the Magellanic Clouds
system (VMC). This survey comprises the Large Magellanic Cloud (LMC),
the Small Magellanic Cloud, the Magellanic Bridge connecting the two
galaxies and two fields in the Magellanic Stream.}{This paper provides
an overview of the VMC survey strategy and presents first science
results. The main goals of the VMC survey are the determination of the
spatially-resolved star-formation history and the three-dimensional
structure of the Magellanic system. The VMC survey is therefore
designed to reach stars as faint as the oldest main sequence turn-off
point and to constrain the mean magnitude of pulsating variables such
as RR Lyrae stars and Cepheids. This paper focuses on observations of
VMC fields in the LMC obtained between November $2009$ and March
$2010$. These observations correspond to a completeness of $7$\% of
the planned LMC fields.}{The VMC data are comprised of multi-epoch
observations which are executed following specific time
constraints. The data were reduced using the VISTA Data Flow System
pipeline with source catalogues, including astrometric and photometric
corrections, produced and made available via the VISTA Science
Archive. The VMC data will be released to the astronomical community
following the European Southern Observatory's Public Survey
policy. The analysis of the data shows that the sensitivity in each
wave band agrees with expectations. Uncertainties and completeness of
the data are also derived.}{The first science results, aimed at
assessing the scientific quality of the VMC data, include an overview
of the distribution of stars in colour-magnitude and colour-colour
diagrams, the detection of planetary nebulae and stellar clusters, and
the $K_\mathrm{s}$ band light-curves of variable stars.}{The VMC
survey represents a tremendous improvement, in spatial resolution and
sensitivity, on previous panoramic observations of the Magellanic
system in the near-infrared, providing a powerful complement to deep
observations at other wavelengths.}


\maketitle

\section{Introduction}
\label{intro}

The cosmological paradigm for the formation and evolution of galaxies
suggests that large structures formed as a sequence of mergers of
smaller objects (White \& Frenk \cite{whi91}). The theoretical
framework relies on cold dark matter simulations and is supported by
high redshift observations (York et al. \cite{yor00}) and by studies
of the cosmic microwave background (Spergel et al. \cite{spe03}), but
the major difficulty is to reproduce the baryonic (stars, gas and
dust) content of the Universe. Therefore, the study of the assembly
process of nearby galaxies via resolved stars is a crucial aspect to
understand how structures in the Universe form and evolve (Tolstoy et
al. \cite{tol09}). In particular, dwarf irregular galaxies are well
suited because their low metallicity and high gas content provide
information about galaxies at an early stage of evolution. The closest
prototypes of interacting dwarf galaxies that offer an excellent
laboratory for this near-field cosmology are the Magellanic Clouds
(MCs).

The Magellanic system is located at a distance of $\sim57$ kpc
(e.g. Cioni et al. \cite{cio00b}) and comprises: the Large Magellanic
Cloud (LMC), the Small Magellanic Cloud (SMC), the Magellanic Bridge
and the Magellanic Stream. The LMC is a dwarf irregular galaxy seen
nearly face-on (e.g. van der Marel \& Cioni \cite{vdm01}) and
sometimes referred to as a late-type spiral galaxy, rich in gas and
actively forming stars. The SMC is a highly inclined dwarf irregular
galaxy also referred to as a dwarf spheroidal galaxy (Zaritsky et
al. \cite{zar00}) with less active star formation. The LMC is probably
just a few kpc thick along the line-of-sight, but the SMC has a more
complex structure that may extend up to 20 kpc along the line-of-sight
(e.g. Westerlund \cite{wes97}, Groenewegen \cite{gro00}, Subramanian
\& Subramaniam \cite{sub09n}). There is a bar embedded in each galaxy
(Subramaniam \& Subramanian \cite{sub09m}, Gonidakis et
al. \cite{god09}). The Magellanic system is metal-poor, the
metallicity is about $1/2$, $1/4$ and $1/10$ that of the Sun for the
LMC, SMC and the Bridge, respectively. The MCs have experienced an
extended star formation history (e.g. Hill et al. \cite{hil00};
Zaritsky et al. \cite{zar02}, \cite{zar04}; Cole et al. \cite{col05};
Pomp\'{e}ia et al. \cite{pom08}; Gallart et al. \cite{gal08}; Carrera
et al. \cite{car08}).

The dynamical interaction between the MCs may be responsible for the
various episodes of star formation (Zaritsky \& Harris \cite{zar04})
and for the creation of the Bridge (Irwin et al. \cite{irw85}, Gordon
et al. \cite{gor09}) which connects the two galaxies and clearly
has young stars associated with it (Irwin \cite{irw91}, Battinelli \& Demers
\cite{bat98}). The Stream appears (to date) as
a purely gaseous feature spanning more than $100$ deg in the Southern
sky (Guhathakurta \& Reitzel \cite{guh98}). A tidal origin of the
Stream from the interaction between the LMC and the Milky Way (MW) has
been ruled out by new proper motion measurements (Kallivayalil et
al. \cite{kal06a}, \cite{kal06b}). Alternative explanations are: ram
pressure (Mastropietro et al. \cite{mas05}) and tidal origin from the
interaction between the LMC and the SMC (Besla et al. \cite{bes10}).

The interaction between the MCs and the MW is representative of the
environmental effects that large galaxies with satellites (low-mass
dwarf galaxies) experience elsewhere in the Universe. This suggests
that the MCs may have entered the Local Group as part of an
association (cf. Tully et al. \cite{tul06}, Moss \cite{mos06}, Knebe
et al. \cite{kne06}) and that, in the future, a minor merger between the
LMC and the MW may occur (cf. Ibata et al. \cite{iba94},
\cite{iba03}). An alternative is that the MCs may be tidal dwarfs expelled during
a previous merger event involving M31 (Yang \& Hammer \cite{yan10}),
although this would imply they have retained little dark matter from
their parent halos (Barnes \& Hernquist \cite{bar92}). At present,
basic assumptions are being challenged: What is the origin of the MCs? 
Do the MCs constitute a binary system and, if so, for how long? Have
the MCs interacted with the MW or are they on their first approach
(Besla et al. \cite{bes07}, D'Onghia \& Lake \cite{don08})? How have
the star-formation histories of the LMC and the SMC been influenced by
interaction? Does the geometry of the system depend on age and
metallicity? How do star clusters form and evolve in the MCs?  What is
the fate of the MCs and will they merge with the MW? Will the Bridge
evolve into a dwarf galaxy (Nishiyama et al. \cite{nis07})? Does the
LMC have an ordinary bar and how does it influence the LMC evolution? 
Or, is the offset bar a separate galaxy being merged into the LMC
disc? Does the LMC have a metal poor old halo? Why is there a
significant difference in structure between the gas and stars in the
SMC? Does the SMC have a bulge?

To answer all these questions we must resolve the stellar populations
and study them in detail. A fundamental step in this direction has been the
many panoramic imaging surveys that have provided multi-wavelength
observations of the Magellanic system. Except for the dedicated
optical MCPS survey (Zaritsky et al. 2002, 2004) information about the
overall population of the MCs has been obtained from surveys with
different original goals, including microlensing optical surveys
(e.g. MACHO - Alcock et al. \cite{alc00}, EROS - Tisserand et
al. \cite{tis07}, and OGLE - Udalski et al. \cite{uda92}), and infrared
sky surveys (e.g. IRAS - Schwering \cite{sch89}, 2MASS - Skrutskie et
al. \cite{skr06}, and DENIS - Cioni et al. \cite{cio00a}). A
continuation of OGLE is still in progress while other large-scale,
near-infrared (near-IR) surveys (IRSF - Kato et al. \cite{kat07})
offer somewhat more sensitive data than 2MASS. The surveys in the
mid-infrared with the {\em Spitzer Space Telescope} (SAGE - Meixner et
al. \cite{mei06}, Bolatto et al. \cite{bol07}, Gordon et
al. \cite{gor09}, Bonanos et al. \cite{bon10}) of the central part of
the galaxies and an optical survey of the outermost regions (Saha et
al. \cite{sah10}) have recently been completed. These surveys have
provided data covering most of the electromagnetic spectrum, but their
common depth is limited to moderately bright giant stars. The
development of the VISual and Infrared Survey Telescope for Astronomy
(VISTA - Emerson \& Sutherland \cite{eme10}) offers a unique
opportunity to acquire near-IR data of unprecedented sensitivity in
the Magellanic system. This is the underlying objective of the near-IR
$Y J K_\mathrm{s}$ VISTA survey of the Magellanic Clouds system (VMC).

This paper is organised as follows. Section \ref{prog} introduces the
VMC survey and then describes the observing strategy and first
observations. Section \ref{data} describes the data reduction steps
for producing images and catalogues for individual observations, while
Sect. \ref{vsa} describes the subsequent stages of reduction for deep
and linked observations and presents the archival procedures. Section
\ref{results} shows results from the first data. Section
\ref{conclusions} concludes this study and the Appendix gives the
coordinates of the VMC fields.

\section{VMC survey}
\label{prog}

The VMC\footnote{http://star.herts.ac.uk/$\sim$mcioni/vmc} is a
uniform and homogeneous survey of the Magellanic system in the near-IR
with VISTA. The main parameters of the survey are summarised in
Tab. \ref{vmcparam}. It is the result of a letter of intent submitted
in $2006$ and a science and management plan approved early in $2008$.
The main science goals of the survey are the determination of the
spatially-resolved star-formation history (SFH) and the
three-dimensional (3D) structure of the Magellanic system. VMC
observations will detect stars encompassing most phases of evolution:
main-sequence stars, subgiants, upper and lower red giant branch (RGB)
stars, red clump stars, RR Lyrae and Cepheid variables, asymptotic
giant branch (AGB) stars, post-AGB stars, planetary nebulae (PNe),
supernova remnants (SNRs), etc. These different populations will help
assess the evolution of age and metallicity within the system. 

The SFH will be recovered from the analysis of colour-magnitude
diagrams (CMDs) and simulations of the observed stellar populations,
accounting for foreground stars and extinction. Kerber et
al. (\cite{ker09}) show a preliminary assessment of the SFH accuracy
that can be expected from VMC data. Modelling near-IR colours bears no
greater uncertainty than optical colours and the near-IR is
particularly sensitive to the colour of the oldest turn-off stars. The
3D geometry will be derived using different density and distance
indicators like the luminosity of red clump stars and the Cepheid and
RR Lyrae period-luminosity relation, and period-luminosity-colour and
Wesenheit relations. These results will complement those based on
2MASS data for the AGB and upper RGB populations of the LMC (van der
Marel \cite{vdm}) and those from optical data of the central regions
of the galaxies that are affected by a higher reddening
(e.g. Subramanian \& Subramaniam \cite{sub10}) and crowding. We will
constrain the epoch of formation of each galactic component by mapping
the extent of different kinds of stars and deriving, using up-to-date
stellar evolutionary models, ages and metallicities. The VMC survey
science addresses many other issues in the field of star and galaxy
formation and evolution, such as: stellar clusters and streams;
extended sources; proper motions; star formation; distance scale;
models of Magellanic system evolution; extinction mapping.

\begin{table*}
\caption{VMC survey parameters.}
\label{vmcparam}
\[
\begin{tabular}{lccc|lccc}
\hline
\noalign{\smallskip}
Filter & $Y$ & $J$ & $K_\mathrm{s}$ &  Filter & $Y$ & $J$ & $K_\mathrm{s}$\\
\hline
\noalign{\smallskip}
Central wavelength ($\mu$m) & 1.02 & 1.25 & 2.15 & Exposure time per epoch (sec) & 800 & 800 & 750 \\
Bandwidth ($\mu$m) & 0.10 & 0.18 & 0.30 & Number of epochs & 3 & 3 & 12 \\ 
Detector Integration Time - DIT (sec) & 20 & 10 & 5 & Total exposure time (sec) & 2400 & 2400 & 9000\\
Number of DITs & 4 & 8 & 15 & Predicted sensitivity per epoch (Vega mag) & 21.3 & 20.8 & 18.9 \\
Number of exposures & 1 & 1 & 1 & Signal-to-noise per epoch at depth required & 5.7 & 5.9 & 2.9 \\
Micro-stepping & 1 & 1 & 1 & Total predicted sensitivity (Vega mag) & 21.9 & 21.4 & 20.3 \\
Number of Jitters & 5 & 5 & 5  & Total signal-to-noise at depth required & 10 & 10 & 10  \\
Paw-prints in tile & 6 & 6 & 6 &  Saturation limit (Vega mag) & 12.9 & 12.7 & 11.4 \\
Pixel size (arcsec) & 0.339 & 0.339 & 0.339 & Area (deg$^2$) & 184 & 184 & 184 \\
System FWHM & 0.51 & 0.51 & 0.51 & Number of tiles & 110 & 110 & 110 \\
\noalign{\smallskip}
\hline
\end{tabular}
\]
Jitter pattern $= Jitter5n$. Tile pattern $= Tile6zz$. 
\end{table*}

\subsection{The VISTA telescope and camera}

VISTA is a new 4m class telescope developed in the United Kingdom (UK). It formed part of the in-kind contribution of the UK to joining the European Southern Observatory (ESO). The telescope has an alt-azimuth mounting and is located just $1.5$ km from the Very Large Telescope (VLT) site. The VISTA infrared camera (VIRCAM) is equipped with an array of $16$ Raytheon detectors with a mean pixel size of $0.339^{\prime\prime}$ and a field of view of $1.65$ deg$^2$. The VIRCAM has a set of broad-band filters: $Z$, $Y$, $J$, $H$ and $K_{\mathrm s}$ and a narrow-band filter at $1.18\, \mu$m. The point spread function of the system is specified to have $50$\% of the light from a point source contained within a circle of diameter $0.51^{\prime\prime}$. The telescope and its camera are described by Emerson et al. (\cite{eme06}) and Dalton et al. (\cite{dal06}), while the performance during commissioning is presented by Emerson et al. (\cite{eme10}). The science verification programmes are summarised by Arnaboldi et al. (\cite{arn10}). 

VISTA is the largest wide-field near-IR imaging telescope in the world
and it is designed to perform survey observations; at present there
are six Public Surveys
underway\footnote{http://www.eso.org/sci/observing/policies/PublicSurveys/\-sciencePublicSurveys.html}. VISTA
observes a continuous area of sky by filling in the gaps between the
detectors using a sequence of six offsets, each by a significant
fraction of a detector. The combined image corresponds to a VISTA {\it
tile} that covers $\sim 1.5$ deg$^2$, while individual offset
positions are named {\it paw-prints} and cover an area of $0.59$
deg$^2$. The resulting VISTA tile observes each part of sky within it
at least twice, except for two edge strips in the extreme `Y'
directions of the array. The tiling process for the VMC survey ensures
that adjacent tiles overlap sufficiently to provide two observations
of these areas as well. The combination of offsets generates small
overlapping areas with exposures as large as six times a single
paw-print. These steps are described in the VISTA user
manual\footnote{http://www.eso.org/sci/facilities/paranal/instruments/vista/doc/}.

\begin{figure}
\resizebox{\hsize}{!}{\includegraphics{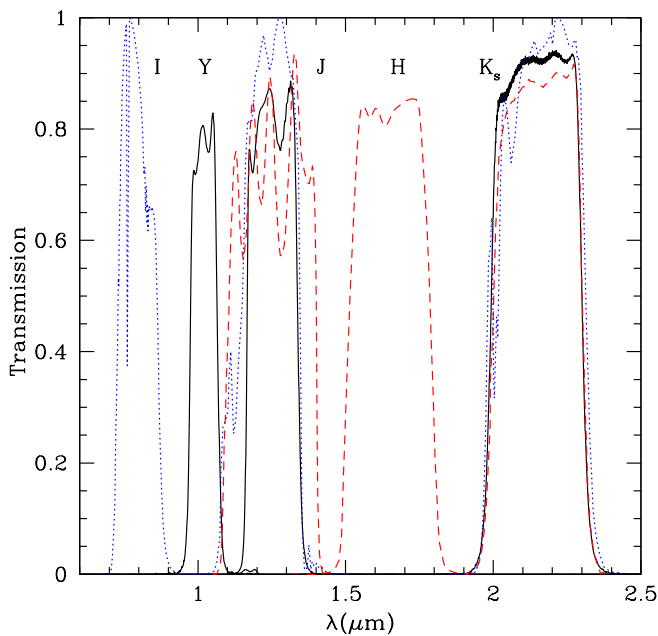}}
\caption{Filter transmission curves for the VMC survey ($YJK_\mathrm{s}$ -- black continuous lines) compared with the transmission of the 2MASS ($JHK_\mathrm{s}$ -- red dashed lines) and DENIS ($IJK_\mathrm{s}$ -- blue dotted lines) surveys.}
\label{filters}
\end{figure}

The VMC survey is a near-IR $YJK_\mathrm{s}$ survey of the Magellanic
system. Compared to previous surveys using the 2MASS and DENIS
telescopes, the choice of VISTA filters for the VMC survey was driven
by the following considerations. A total of three filters were chosen
to be able to analyse colour-colour diagrams. The availability of a
large colour spacing to allow for a good characterisation of the
subgiant branch population to derive the SFH. The relation between
$J-H$ and $J-K_\mathrm{s}$ is quite linear for MC giants, so
observations in the $H$ band would have provided very little
information and priority was given to a bluer filter. The $J$ band was
preferred over $H$ as it also suffers less from atmospheric effects,
and provides comparative observations with respect to previous
surveys.  The choice of $Y$ over $Z$ was motivated by a reduced
confusion limit.  The $K_\mathrm{s}$ filter was required to
determine the average magnitude of variable stars that, at this
wavelength, obey a clear period-magnitude relation that is unaffected
by other stellar parameters that degrade the relation at bluer
wavelengths (Sect. \ref{variables}). A comparison among the filter
transmission curves for the VMC, 2MASS and DENIS observations is shown
in Fig. \ref{filters}. The exposure time for the VMC survey
(Tab. \ref{vmcparam}) is designed to meet the two key scientific
objectives: the SFH and the 3D geometry. An accurate determination of
the SFH requires CMDs reaching the oldest main sequence turn-off, to
allow for sampling of different stellar populations. The investigation
of RR Lyrae stars and Cepheids requires monitoring observations across
specific time intervals. The stacking of the observations that are
needed for deriving the mean magnitude of variable stars in the
$K_\mathrm{s}$ band meets also the depth requirement for the SFH. The
split of epochs in the $Y$ and $J$ filters is instead purely driven by
scheduling requirements.

\subsection{VMC area coverage}
Observing the entire Magellanic system extending over hundreds of kpc
(the Stream covers half the sky) is a daunting task. The VMC survey
concentrates, therefore, on a moderately extended area ($\sim$180
deg$^2$) that includes the classical diameter limit at $B\approx25$
mag arcsec$^{-2}$ for both galaxies (Bothun \& Thompson \cite{bot88})
as well as major features traced by the distribution of stars
(e.g. Irwin \cite{irw91}, Bica et al. \cite{bic08}) and H\,{\footnotesize I\normalsize}
gas (e.g. Staveley-Smith et al. \cite{sta03}, Hatzidimitriou et
al. \cite{hat05}, Muller et al. \cite{mul03}), the Bridge and two
fields at specific locations in the Stream.

\begin{figure*}
\resizebox{\hsize}{!}{\includegraphics[angle=270]{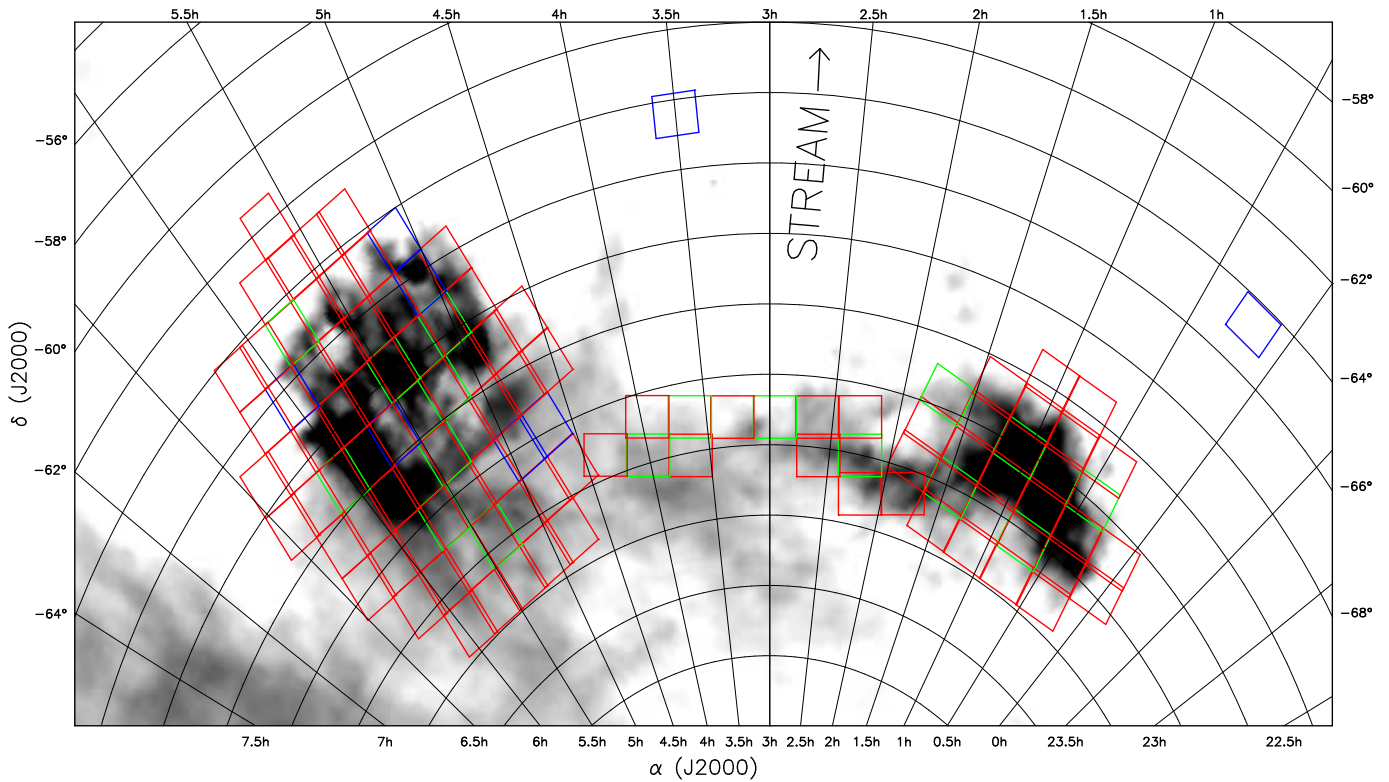}}
\caption{Magellanic system area tiled for VMC observations. The 
underlying image shows the HI distribution (McClure-Griffiths et
al. \cite{mcc09}). VISTA tiles are colour coded as follows. Blue
rectangles represent tiles for which observations have started during
the dry-runs and P85, green rectangles are for tiles with observations
in P86, and red tiles are observations that will not being before
P87.}
\label{tiles}
\end{figure*}

The LMC area ($116$ deg$^2$) is covered by $68$ tiles, while $27$
tiles cover the SMC ($45$ deg$^2$) and $13$ cover the Bridge ($20$
deg$^2$), see Fig. \ref{tiles}. Additionally, $2$ tiles ($3$ deg$^2$)
are positioned in the Stream, one approximately to the North of the
centre of the Bridge (corresponding to a dense area of gas) and the other, at
similar right ascension, to the North of the SMC, corresponding to a
dense area of stars following the simulations by Mastropietro
(\cite{mas09}). VMC tile centre coordinates are given in the
Appendix. Each tile is identified by two numbers: the first number
indicates the row and the second the column of the position of the
tile in the mosaic that covers the system. Note that a separate tiling
pattern has been defined for each region (Figs. \ref{tileslmc}, \ref{tilessmc},
\ref{tilesbridge}). Row numbers increase from South to North and
column numbers increase from West to East.

Tiles covering the LMC were oriented at a position angle of $+90$
deg. The default orientation (position angle $=0$ deg) points the `Y'
axis to the North and the `X' axis to the West. The position angle is
defined to increase from North to East. This represents the best
compromise between minimum number of tiles and maximum area,
increasing the efficiency of the survey. The overlap between the
doubly-covered sky areas in adjacent tiles corresponds to
$60^{\prime\prime}$ in both `X' and `Y' directions. The LMC mosaic was
created using the Survey Area Definition Tool (SADT -- Arnaboldi et
al. \cite{arn08}). A geodesic rectangle centred at
$\alpha=05$:$35$:$50$, $\delta=-69$:$27$:$12$ (J2000), with width
$=11.8$ deg and height $=15.9$ deg, was created as the basis of the
tiling process. Outer tiles were removed leading to the pattern shown
in Fig. \ref{tiles} (see also Fig. \ref{tileslmc}). The area covered
by the tiles was checked against the distribution of stellar
associations, carbon stars and other stellar objects using Aladin
(Bonnarel et al. \cite{bon00}). The centre of the rectangle was
adjusted to include the 30 Doradus nebulosity within a single tile
and, similarly, the field that the future space mission
Gaia\footnote{http://sci.esa.int/science-e/www/area/index.cfm?fareaid=26}
will repeatedly observe for calibration.
%Note that at the location of the Magellanic system the tiles do not follow the lines of constant declination and this implies a larger overlap the closer tiles are to the South Pole.

In the process of creating the mosaic, SADT requires as input the observing parameters that are associated to small (i.e. jittering) and large (i.e. mosaicking) displacements in the tile position. For the VMC survey the maximum jitter was set to $15^{\prime\prime}$, the {\it backtrackStep} to $100$ and the tiling algorithm to $Tile6zz$ (these parameters are described in the SADT user manual). Guide stars were assigned automatically to each tile using the GSC-2 reference catalogue (Lasker et al. \cite{las08}). This process may result in shifting the tile centre in case an insufficient number of reference stars is available, but this was not the case for LMC tiles. 

Tiles covering the SMC region were placed at a position angle of $0$ deg. Keeping the wide tile-edge approximately along the right ascension direction produces maximum coverage for a minimum number of tiles. This also implies centring the geodesic rectangle at $\alpha = 00$:$50$:$00$,  $\delta =-73$:$00$:$00$ (J2000), with width $=8.0$ deg and height $=8.0$ deg. This rectangle represents the basis of the tiling process and outer tiles were subsequently removed leading to the pattern shown in Fig. \ref{tiles} (see also Fig. \ref{tilessmc}). As before, the area covered by tiles was checked against the distribution of different stellar objects.  The position of the centre of the rectangle was tuned to match the area that will be observed in the optical domain by the VST\footnote{http://vstportal.oacn.inaf.it/} (VLT Survey Telescope) as part of the STEP survey (P.I. Ripepi; Capaccioli et al. \cite{cap05}) and to provide sufficient overlap for a consistent calibration with the Bridge area. 

Previous observations of the Magellanic Bridge by Battinelli \& Demers (\cite{bat92}) and Harris (\cite{har07}) covered fields departing from the LMC and tracing arcs at different declinations with only a few fields between the two. In order to maximise the total population of stars that the VMC survey will detect the 2MASS, DENIS and SuperCOSMOS\footnote{http://surveys.roe.ac.uk/ssa/index.html} databases were explored for stellar densities and compared with the previous observations. VMC tiles were then positioned to overlap with the area that provides a good sampling of the stellar population while also following a continuous pattern. Following a similar procedure for the LMC and SMC areas, a geodesic rectangle was drawn with centre at $\alpha = 03$:$00$:$00$, $\delta=-74$:$30$:$00$ (J2000), with width $13.5$ deg and height $3$ deg. Outer tiles were subsequently removed leading to the pattern shown in Fig. \ref{tiles} (see also Fig. \ref{tilesbridge}).

The two tiles positioned in the Stream region were prepared using the same parameters as for LMC tiles, except for the position angle set to $0$ deg. Two geodesic rectangles centred on the two fields were defined, each with a default size equal to a single tile.

\subsection{VMC observations}

The VMC data described here were mostly obtained during Science
Verification observations ($15-31$ October $2009$) and the so-called
dry-run period ($1$ November $2009$ -- $31$ March $2010$) when VISTA
was tested and survey operations were still being defined. A small
amount of data was also taken during ESO Period 85 ($1$ April
$2010-30$ September $2010$). The bulk of the VMC observations will be
carried out during the odd--numbered ESO periods starting in October
every year and ending in March the following year because of the
seasonal observability of the Magellanic system.

\begin{table}
\caption{VMC required weather conditions.}
\label{skyconditions}
\[
\begin{array}{ccc}
\hline
\noalign{\smallskip}
\mathrm{Band} & \mathrm{Seeing} & \mathrm{Seeing} \\
                            & \mathrm{(arcsec)} & \mathrm{(arcsec)} \\
                            & \mathrm{uncrowded} & \mathrm{crowded} \\
\hline
\noalign{\smallskip}
Y                         &  1.2 & 1.0 \\ 
J                          & 1.1 & 0.9 \\  
K_\mathrm{s}   & 1.0 & 0.8 \\        
\noalign{\smallskip}                 
\hline
\noalign{\smallskip}
 \mathrm{Field} & \mathrm{Tile\,row} & \mathrm{Airmass} \\
 \hline
\noalign{\smallskip}
\mathrm{LMC} & 2 , 3, 4 & 1.7 \\
\mathrm{LMC} & 5, 6, 7 & 1.6 \\
\mathrm{LMC} & 8, 9, 10, 11& 1.5 \\
\mathrm{Stream}     & 1,2 & 1.5 \\
\mathrm{SMC} & 2 , 3, 4, 5 & 1.7 \\
\mathrm{SMC} & 6, 7 & 1.6 \\
\mathrm{Bridge} & 1, 2, 3 & 1.7 \\
\noalign{\smallskip}
\hline
\end{array}
\]
Moon distance $>80$ deg. Sky transparency $=$ variable, thin cirrus.
\end{table}

Observations of the VISTA Public Survey are obtained in service mode by ESO staff. This guarantees efficiency of operations and a high level of data homogeneity. The requested observing conditions for the VMC survey are summarised in Table \ref{skyconditions}. A dozen tiles, centred on the most crowded regions i.e. 30 Dor and the central regions of both LMC and SMC, have more stringent seeing conditions. This is necessary to prevent confusion in the bluest bands, while $K_\mathrm{s}$ band observations will not be limited by confusion for a seeing $\le 0.9^{\prime\prime}$. The best FWHMs in the VISTA images are $0.6-0.7^{\prime\prime}$ and undersampling, with respect to a pixel size of $0.334^{\prime\prime}$,  is not a  cause of concern in the data treatment.

The Magellanic system never rises above $50^\circ$ from the horizon. Therefore, a compromise had to be made between observing at reasonable airmass and achieving continuous observability over about five months for the monitoring process. The maximum airmass constraints were optimised as a function of the tile declination, as shown in Tab. \ref{skyconditions}. A violation of any observing constraint by $10$\% is still considered as if observations were obtained within specifications. It is also assumed that the airmass constraint may be violated by more than $10\%$ provided the seeing constraint is not. 

Table \ref{vmcparam} describes the main parameters of the VMC observations. The total exposure time is calculated as follows: (number of epochs) $\times \,2\, \times$ (number of jitters) $\times$ (number of DITs) $\times$ DIT. The factor of two comes from the tiling pattern, during which most points of the sky are observed twice (on average). The exceptions are the tile edges, observed once, and some areas of extra overlap among the detectors that are observed four or six times. For example, for the $K_\mathrm{s}$ band the total exposure time is: $12 \times 2 \times 5 \times 15 \times 5=9000$ sec.
   
The observations follow the nesting sequence FPJME (see VISTA User
Manual\footnote{http://www.eso.org/sci/facilities/paranal/instruments/vista/doc/}). This
sequence first sets a filter, then obtains images at all jittered
positions of the first paw-print, before moving to the next paw-print
and taking all the jittered images at that position, and so on, until
all six paw-prints that form the tile are completed. The jittern
pattern was {\it jitter5n} and the tile pattern was {\it Tile6zz}.

The progress of the VMC survey is shown in  Fig. \ref{tiles} and detailed in Tab. \ref{vmcprogress}. The reader is referred to the VMC$^1$ public web page for following up the survey progress beyond that described in this paper.

\begin{table}
\caption{VMC survey progress.}
\label{vmcprogress}
\[
\begin{tabular}{lrrr}
\hline
\noalign{\smallskip}
Description & VMC & LMC & LMC \\
    & all & all & season I \\
\hline
\noalign{\smallskip}
Total number of tiles & 110 & 68 & 6 \\
Total number of epochs & 1980 & 1224 & 108 \\
Total number of $Y$-band epochs & 330 & 204 & 18 \\
Total number of $J$-band epochs & 330 & 204 & 18 \\
Total number of $K_\mathrm{s}$-band epochs & 1320  & 816 & 72 \\ 
Observed number of $Y$-band epochs & 18 & 18 & 18 \\
Observed number of $J$-band epochs & 18 & 18 & 18 \\
Observed number of $K_\mathrm{s}$-band epochs & 51 & 51 & 51 \\
Total number of observed epochs & 87 & 87 & 87 \\
Completion in the $Y$-band & 5.4\% & 8.8\% & 100\% \\ 
Completion in the $J$-band & 5.4\% & 8.8\% & 100\% \\
Completion in the $K_\mathrm{s}$-band & 3.9\% & 6.2\% & 71\% \\
Total completion & 4.4\% & 7.1\% & 80\% \\
\noalign{\smallskip}
\hline
\end{tabular}
\]
\end{table}

\subsubsection{Science Verification}

The observation of a single paw-print in the $YJK_\mathrm{s}$ filters was requested during the instrument Science Verification (SV) time. The main goal was to verify the observing strategy and to test the saturation limits for the adopted DITs of $6$, $10$ and $20$ sec in the $K_\mathrm{s}$, $J$ and $Y$ bands, respectively.  The observations were carried out on the night of $17$ October $2009$, but the data were only available after the dry-runs had started. 

The SV data indicate that the saturation limit varies from one detector to another. On average, it is $11.6$ mag for the $6$ sec $K_\mathrm{s}$ band exposure, with seeing of $0.94^{\prime\prime}$. The linearity is mild at $11.6$ mag, but becomes severe for point sources brighter than $10.7$ mag. These limiting magnitudes are somewhat fainter than predicted.  Prior to the beginning of the dry-runs it was decided to reduce the $K_\mathrm{s}$ band DIT down to $5$ sec, which improves the photometry of the bright stars and maximises the range of overlapping magnitudes with 2MASS. The observing strategy was then established and no modification was needed for the observations that had already started.

Other SV observations were obtained on $28$ Nov $2009$ in the SMC. These are $K_\mathrm{s}$ band observations of one tile, with a six paw-print mosaic and DIT$=10$ sec, with the purpose of checking the sky-subtraction procedures. Images were obtained to test an off-sky algorithm which is not applicable to VMC data. In addition, some 2MASS touchstone fields that are observed for photometric checks are located in the VMC survey area, providing extra, albeit shallow (DIT$=5$ sec), multi-epoch data in all VIRCAM filters.

\subsubsection{Dry-runs during $1$ November $2009$ -- $14$ February $2010$}

One hundred hours of VMC observations were submitted for execution during the $1$ November $2009 - 14$ February $2010$ dry-run period and $70$\% were executed. Tables \ref{otab} and \ref{ktab} summarise the observations where each date corresponds to the observing date of one observing block (OB). Each VMC OB includes only one filter. These observations were prepared using a new version of the Phase II Proposal Preparation\footnote{http://www.eso.org/sci/observing/phase2/P2PP/P2PPSurveys.html} tool especially revised for Public Surveys with the VISTA and VST telescopes. It allows the user to have more control over the survey execution with new high-level tools called {\it scheduling containers} (Arnaboldi et al. \cite{arn08}). Three types of containers were available: {\it concatenations} (grouping together OBs for back-to-back execution), {\it time-links} (imposing time constraints for the execution of OBs, including execution of OBs in a user-defined sequence), and {\it groups} (that improve the prioritisation of OBs to ensure the completion of one set of OBs before another set is started). 

Six VMC fields were observed in the LMC (Fig. \ref{tiles}). One field covers the famous 30 Dor region ($6\_6$), one field corresponds to the South Ecliptic Pole (SEP) region that Gaia will observe repeatedly for calibration ($8\_8$), a pair of fields are located in the Northern outer part of the LMC disc ($8\_3$ and $9\_3$, overlapping in declination) while the remaining two fields overlap in right ascension and are located towards the Bridge ($4\_2$ and $4\_3$). The 30 Dor field was chosen because it is well studied and represents a crowded region of the LMC rich in stars and gas. The Gaia field was chosen for its importance to the Gaia mission; early analysis and release of these data will benefit the astronomical community in general. The two fields in the Northern disc were chosen to sample an uncrowded and external area of the galaxy. Those towards the Bridge satisfy a similar criterion but sample a different region of the LMC disc. Overall the six fields were also chosen to reflect different coordinates and observing conditions.

Table \ref{otab} shows the observations of concatenations and groups for each LMC field. Groups represent observations in the $Y$ and $J$ bands that are not linked by time, i.e. for a given field they can be executed on any night including on the same night. They are labelled $Y1$, $J1$, $Y2$ and $J2$.  Concatenated observations are preceded by the letter $C$ and the integration time per filter is shorter than for a single VMC epoch at that filter (Tab. \ref{vmcparam}), but the different filters are observed back-to-back. The time limit for a concatenation, during this period, was $1.5$ hours.  Two concatenations were, therefore, necessary for recovering one VMC epoch per filter. An important feature of concatenations is that the sky condition requirements are met during its entire duration. This is not always the case and then the concatenation has to be repeated. This explains why in Tab. \ref{otab} there are additional observations of OBs that are part of concatenations but without observations in all three filters. These additional observations will be used to investigate source variability and to produce deep stacks (Sect. \ref{vsa}) if the observing condition criteria (Tab. \ref{skyconditions}) were met.

Table \ref{ktab} shows time-link sequences where $11$ individual
$K_\mathrm{s}$ band OBs are associated with each LMC field. Each
sequence can start at any time, independent of the execution of
observations listed in Table \ref{otab}, i.e. a $TK1$ OB can be
observed on the same night as an OB in the $Y$, $J$ or $K_\mathrm{s}$
filter taken as part of a concatenation or group. Once started, the
next observation of a $TKn$ OB, in a time-link sequence, is obtained
at intervals equal or larger than: $1$, $3$, $5$ and $7$ days for
epochs $2$ to $5$, respectively, and thereafter at least $17$ days
from each previous observation (epochs $6$ to $11$).  Some OBs in the
time-link sequence were also repeated.

\begin{table*}
\caption{LMC epochs: concatenations (C) and $YJ$ bands.}
\label{otab}
\[
\begin{array}{l|rrr|rrr|rr|rr}
\hline
\noalign{\smallskip}
\mathrm{Tile} & CK1 & CJ1 & CY1 & CK2 & CJ2 & CY2 & Y1 & J1 & Y2 & J2\\
\hline
\noalign{\smallskip}

4\_2 & 20.12.09 & 20.12.09^b & 20.12.09^b & 15.01.10 & 15.01.10 & 15.01.10 & 15.12.09 &  14.11.09 & 16.12.09 & 28.11.09 \\
         & & & & 4.01.10 & 4.01.10 & 4.01.10 & & & 15.12.09 & \\
4\_3 & 23.12.09 & 23.12.09^a & 23.12.09 & 16.01.10^b & 16.01.10 & 16.01.10 &  9.12.09 & 18.11.09 & 10.12.09 & 28.11.09 \\
         & 22.12.09 & 22.12.09 & 22.12.09 & & & & & & & \\
6\_6 &  5.11.09 &  5.11.09 &  5.11.09 & 20.11.09 & 20.11.09 & 20.11.09 &  4.11.09 &  4.11.09 &  8.11.09 & 8.11.09  \\
         &                &                 &                 &  26.11.09 & 9.11.09  &          &          &          &          &          \\
8\_3 & 26.11.09 & 26.11.09 & 26.11.09 & 22.11.09 & 22.11.09 & 22.11.09 & 19.11.09 & 15.11.09 & 21.11.09 & 26.11.09 \\
8\_8 & 21.11.09 & 21.11.09 & 21.11.09 & 20.11.09 & 20.11.09 & 20.11.09 &  5.11.09 &  9.11.09 &  9.11.09 & 11.11.09\\
         & 26.11.09 & 26.11.09 &          &          &          &          & &  6.11.09^b & 7.11.09^b &         \\
9\_3 & 21.11.09 & 21.11.09 & 21.11.09 & 22.11.09 & 22.11.09 & 22.11.09^b & 30.11.09 & 1.12.09 & 1.12.09  &  2.12.09 \\

\noalign{\smallskip}
\hline
\end{array}
\]
%$^a$ less than six paw-prints available

$^a$ reduced number of jitters and/or paw-prints

$^b$ seeing and/or ellipticity too high

\end{table*}

\begin{table*}
\caption{LMC epochs: time linked (T) $K_\mathrm{s}$-band monitoring.}
\label{ktab}
\[
\begin{array}{l|rrrrrrrrrrr}
\hline
\noalign{\smallskip}
\mathrm{Tile} & TK1 & TK2 & TK3 & TK4 & TK5 & TK6 & TK7 &TK8 & TK9 & TK10 & TK11\\
\hline
\noalign{\smallskip}

4\_2 & 14.12.09 & 17.12.09 & 5.01.10 & 14.01.10 & 22.01.10 & & & & & & \\
4\_3 & 10.12.09 & 17.12.09 & 21.12.09 & 27.12.09 & 18.01.10 & & & & & & \\
6\_6 &  8.11.09 & 12.11.09 & 17.11.09 & 29.11.09 &  7.12.09 & 26.12.09 & 13.01.10^a & 31.01.10 & 19.02.10 & 11.03.10 & 10.11.10\\ 
         &  6.11.09^a & & & & & & & & & & \\
8\_3 &  3.12.09 &  6.12.09 & 22.12.09 & 28.12.09 & 14.01.10 & 31.01.10 & 22.02.10 & 14.11.10^b& & & \\
8\_8 & 14.11.09 & 19.11.09 & 25.11.09 & 30.11.09 &  7.12.09^b & 25.12.09 & 14.01.10 & 31.01.10 & 19.02.10 & 08.03.10 & 26.11.10\\
8\_8 & & & & & & & & & & & 16.11.10^b\\
9\_3 &  4.12.09 &  9.12.09 & 19.12.09 & 23.01.10 & 31.01.10 & 24.02.10 & & & & & \\
 & & & & 16.01.10^b & & 20.02.10^b & & & & & \\
\noalign{\smallskip}
\hline
\end{array}
\]
$^a$ reduced number of jitters and/or paw-prints

$^b$ seeing and/or ellipticity too high

\end{table*}

\subsubsection{Dry-runs during $15$ February $2010$ -- $31$ March $2010$ and ESO Period 85}

The remaining dry-run period until the end of March $2010$ and the subsequent ESO survey period (P85) until the end of September $2010$ were not ideally suited for observations of the Magellanic system, except for short windows at the beginning and the end of the period. The VMC survey was assigned $30$ hours during this period. Seven hours were scheduled to progress on the $K_\mathrm{s}$ monitoring of four of the six LMC fields started during the first dry-run period. The remaining hours were for observations of the fields in the Stream. The overall observation of the LMC fields during the dry-runs and P85 bring the completion rate of the six fields to $80$\%.

The strategy for the fields in the Stream was changed because concatenations were limited to a total execution of time of $1$ hour. This limit imposes a further subdivision of the VMC epochs that may increase the observing overheads. We have decided to adopt a different procedure that keeps the overheads more or less constant and does not interfere with the requirement of observing different filters as close in time as possible. In the new approach there are $3$ concatenations per field and each one contains only two filters: $YJ$, $YK_\mathrm{s}$ and $JK_\mathrm{s}$. The sum of the integration times per filter is the same as that from the two previous concatenations and it is also the same as a VMC epoch. The first observations of the Stream will be described in a subsequent paper of this series.

\section{Data Reduction}
\label{data}

The raw VISTA images acquired for the VMC survey were reduced by the VISTA Data Flow System (VDFS) pipeline at the Cambridge Astronomical Survey Unit (CASU\footnote{http://casu.ast.cam.ac.uk/surveys-projects/vista}). The VDFS pipeline is specifically designed for reducing VISTA data (Irwin et al. \cite{irw04}) and is used to process up to 250 GB/night of data. The pipeline is a modular design allowing straightforward addition or removal of processing stages. The VMC data are reduced together with other VISTA data on a weekly basis. Prior to this science reduction the data are checked at the observatory site (ESO Chile) using a simplified version of the VDFS pipeline and library calibrations. The data are subsequently checked at ESO in Garching for monitoring the instrument performance and to feed updated information back to the observatory.

The most relevant VDFS steps for the reduction of VMC survey data are as follows: 
\vspace{-0.2cm}
\begin{itemize}
\item reset, dark, linearity, flat and de-striping (removing horizontal stripes in the background) correction; 
\item sky background correction (tracking and homogenisation during image stacking and mosaicking); 
\item jittered and paw-print stacking;
\item point source extraction; 
\item astrometric and photometric calibration (the latter put in an internally uniform system); 
\item bad pixel handling, propagation of uncertainties and effective exposure times by use of confidence maps; 
\item nightly extinction measurements.
\end{itemize}

The different observational uncertainties are propagated during the data processing, to give the users a clear picture of the final data quality. Various quality control parameters are calculated during the data reduction to monitor the data, and to evaluate both the observing conditions (in retrospect) during the observation, and the individual data reduction steps. Among them are: the zero-point to measure the atmospheric extinction, the FWHM to measure the seeing, the ellipticity to evaluate the quality of the guiding and active optics correction, the sky level to estimate the background level and its variations, etc.. The processing history is recorded directly in FITS headers.

A tile image is produced by combining $96$ different images ($16$ detector images per each of $6$ paw-prints). Their sky level and individual paw-print astrometric and photometric distortion are adjusted in the drizzling (combination) process. Tile catalogues are produced following the application of a nebulosity filter to the paw-prints in order to remove diffuse varying background on scales of $30^{\prime\prime}$ or larger (Irwin \cite{irw10}). This method has shown that the detection of objects and their characterisation (astrometry, photometry and morphological classification) are considerably improved. 

The first VMC reduced images and catalogues, corresponding to observations obtained during November $2009$, were received in January $2010$. These data were reduced with version $0.6$ (v$0.6$) of the VDFS pipeline and refer to individual paw-prints, not yet combined into tiles. This means that in practise, for example, the observation of a $Y$-band OB of a given field has $6$ associated images, $6$ catalogues and $6$ confidence maps. Each image and catalogue are delivered in Rice compressed format and are multi-extension FITS files containing each the information for all of the $16$ detectors covering the VIRCAM field-of-view. In March $2010$ we received more data processed with an upgraded version of the pipeline (v$0.8$). These data include all observations obtained until the end of January $2010$. The processed observations obtained after this date and until the end of the Magellanic season (March $2010$) were received in April $2010$. Following quality inspection a few images were re-reduced (v$0.9$) to fix some specific problems with a small subset of the data. The v$0.8$ data were ingested into the VISTA Science Archive (VSA) by June $2010$ (Sect. \ref{vsa}). They include astrometry and photometry in single-band, band-merged and epoch-merged tables as well as deep stacks. In October $2010$ v$1.0$ data have become available and the main difference from previous releases is that they include tile images and catalogues.

\begin{figure*}
\resizebox{18cm}{22cm}{\includegraphics{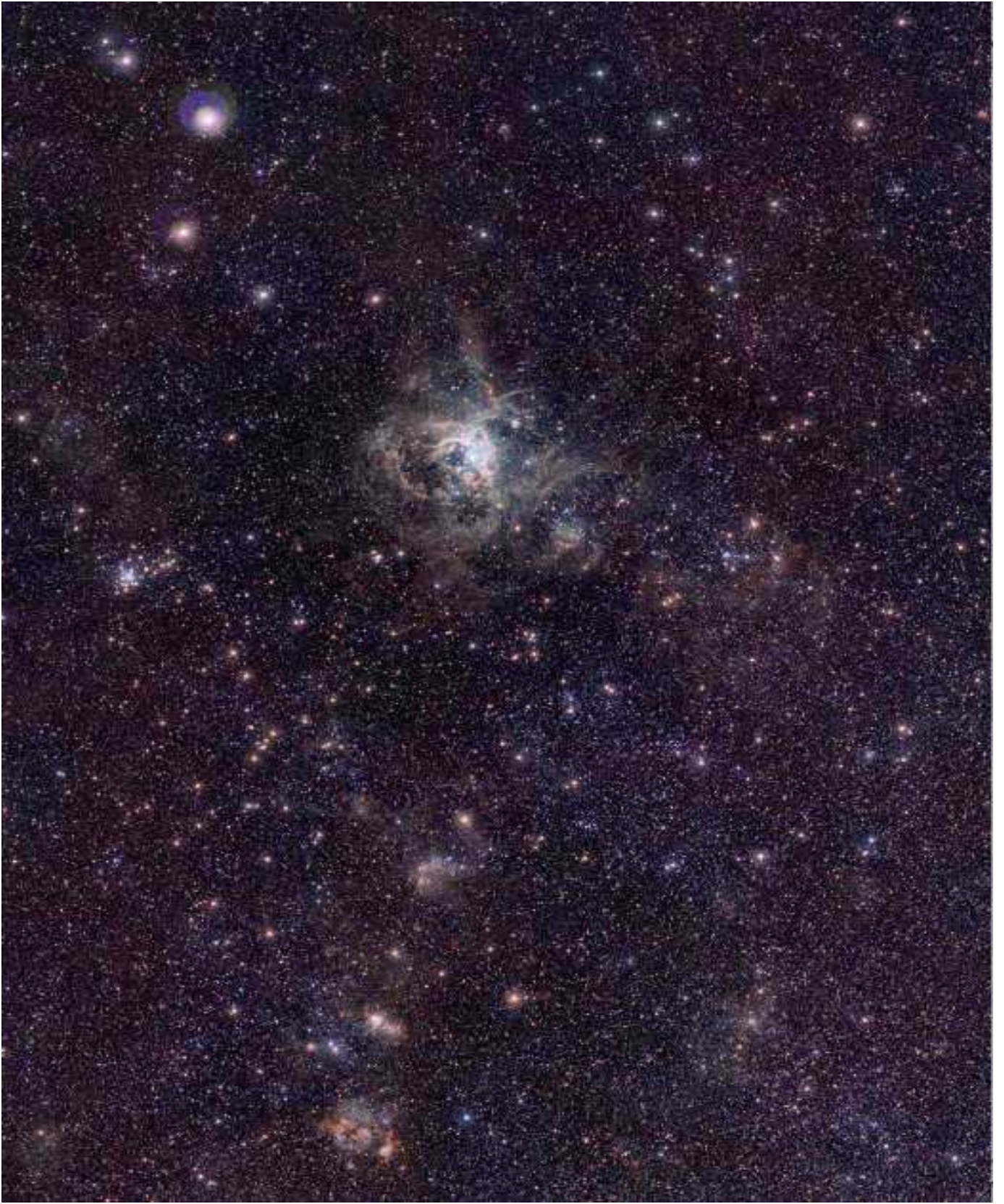}}
\caption{Most of VMC tile of LMC field 6\_6. This is a colour composite image where $Y$ is shown in blue, $J$ in green and $K_\mathrm{s}$ in red. East is to the left and North at the top. The 30 Doradus star forming region is visible together with other smaller regions towards the South as well as stellar clusters and the field population. For more details and a high resolution image refer to http://www.eso.org/public/news/eso1033/.}
\label{30dor}
\end{figure*}

Figure \ref{30dor} shows most of a tile of the 6\_6 LMC field including the star forming region 30 Dor. This image was produced for an ESO press release\footnote{http://www.eso.org/public/news/eso1033/} where other zoomed-in images are also available. The exposure time in the three wave bands was $2400$ sec in $Y$, $2800$ sec in $J$ and $4850$ sec in $K_\mathrm{s}$.

\subsection{Linearity and background}
\label{saturation}

The individual VISTA detectors have different non-linearity and
saturation properties and these properties may also vary across a
given detector. The detector system is non-linear and linearity
corrections are applied at the pixel level during the initial image
processing stages. The saturation levels are stored in the image
header keywords and, together with the peak flux derived from the
photometry of the observed sources, it is possible to establish a
correction that enables recovery of stars up to a few 
magnitudes brighter than the saturation limit (Irwin
\cite{irw09}). This correction is not implemented by VDFS prior to the
production of VMC catalogues but it is applied at a further stage of
the data processing (Sect. \ref{vsa}). The expected saturation values
listed in Tab. \ref{vmcparam} are sensitive to seeing variations. In
fact, the 30 Dor region was observed under very good
conditions, and saturation appears at fainter magnitudes.

%Non-linearity at the $1-2$\% level for flux values $<1000$ counts is present in the data. This small effect is not corrected in a standard linearity sequence because the count level is too low. The most non-linear detectors are \#13 and \#16.

The application of a nebulosity filter to the paw-print images prior to the construction of a tile image (Irwin \cite{irw10}), may influence the recovery of the magnitude of stars close to the saturation limit. This would, however, only affect heavily saturated stars, i.e. those where a significant halo with a diameter comparable to  that of the filter size ($\sim 30^{\prime\prime}$) is visible.

The sky background for VMC observations is estimated, for each
paw-print, from all paw-prints observed for one tile in a given band
and at a given time. This method, referred to as `tilesky', has shown
very good results even for the 30 Dor tile where there is a
substantial emission from the nebula.

Persistence effects due to bright stars are usually automatically removed by the VDFS pipeline when the observations, like for VMC, follow the FPJME sequence (see VISTA User Manual$^6$). Adding up all VMC images for a given field and filter does not produce any noticeable effect due to persistence.

The moon has a negligible effect on the VMC background because it is
always $>$80 deg away from any of the fields. This results in low
contamination, even in the $Y$ band which is most susceptible to the
lunar contamination. The major absorption is caused by water vapour and
carbon dioxide in the atmosphere. At the VMC filters the background
will also be dominated by non-thermal aurora emission, OH and O$_2$
lines especially in observations obtained $1.5-2$ hours after
twilight.

\subsection{Astrometry}

Astrometry is based on positions of the many 2MASS sources within each
detector. The astrometric calibration of a paw-print is encoded in the
FITS image headers using the Zenithal Polynomial projection (ZPN)
while a tile refers to a single tangent plane World Coordinate System
(WCS) image (Calabretta \& Greisen \cite{cal02}). The median
astrometric root-mean-square is $80$ mas and is dominated by 2MASS
uncertainties. Residual systematic distortions across the VISTA
field-of-view are at the $\sim$25 mas level and can be further
improved, if required, by directly characterising the distortion
pattern.  In a dithered sequence the detectors are rotated slightly to
maintain the position angle on the sky; a comparison between identical
VISTA observations shows a residual rotation of $\sim$0.5 pix.

\begin{figure}
\resizebox{\hsize}{!}{\includegraphics{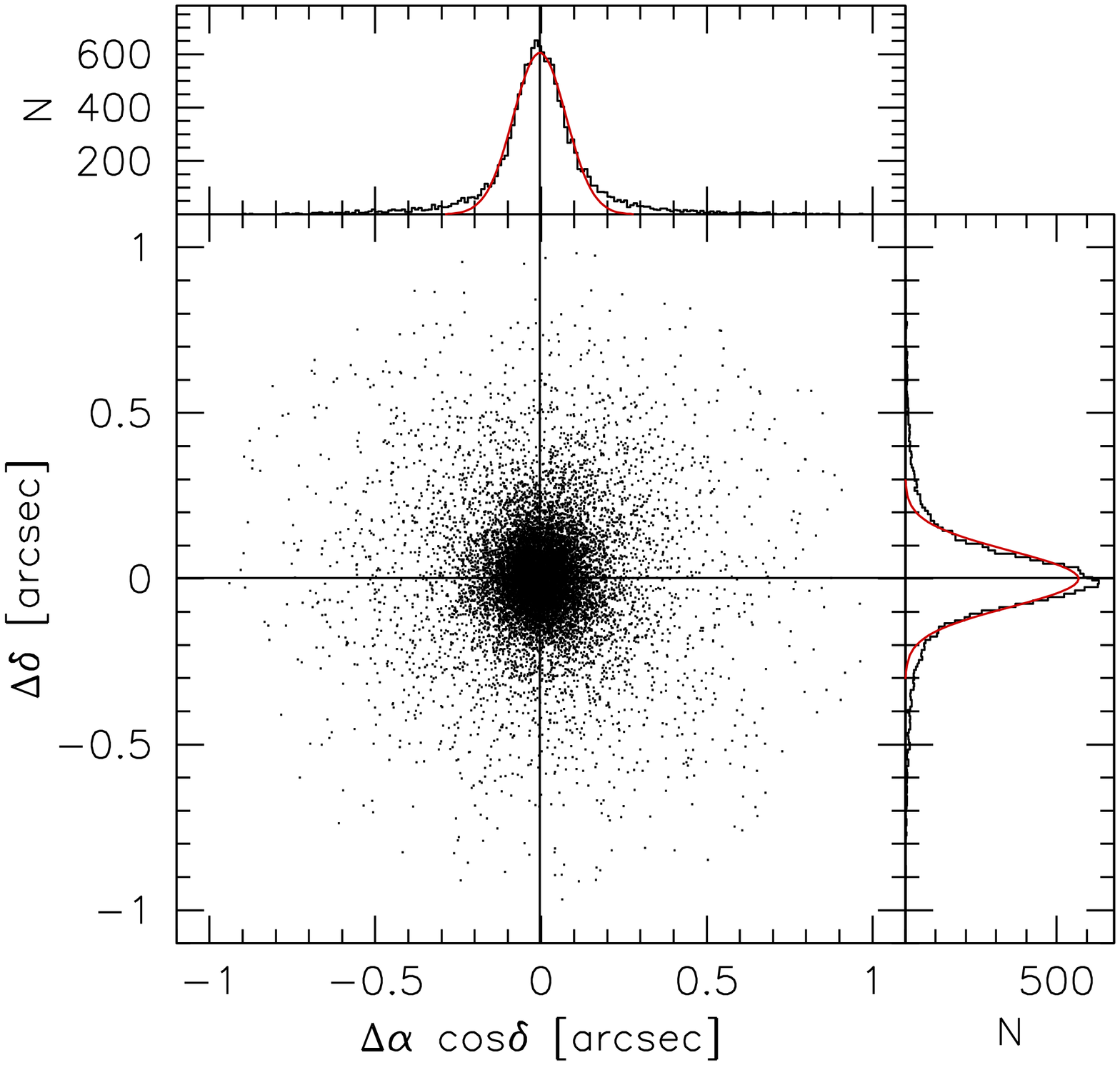}}
\caption{Comparison between VISTA and 2MASS astrometry in the 8\_8 field. Histograms have bins of $0.01^{\prime\prime}$ in size. The best fitting gaussians are indicated and correspond to $\sigma$ of $0.080^{\prime\prime}$ and $0.085^{\prime\prime}$ for $\Delta\alpha$ and $\Delta\delta$, respectively.}
\label{astrometry}
\end{figure}

Figure \ref{astrometry} shows a comparison between the right ascension ($\alpha$) and declination ($\delta$) coordinates of stars that are in common between VISTA and 2MASS in the VMC field 8\_8. The excellent match shows the quality of the astrometry. A systematic shift is perhaps present at the level of  $\sim 0.01^{\prime\prime}$ in both axes. This accuracy is perfectly adequate for cross-correlation studies between external catalogues and VMC. The relative accuracy within VMC data is higher and a more detailed investigation of the suitability of VMC data for studies of, for example, proper motions will be addressed elsewhere.

\subsection{Photometry}

The photometric calibration relies on the observation of stars from the 2MASS catalogue with magnitudes in the range $12-14$ mag in all bands. The procedure is similar to that adopted to calibrate data from the Wide-Field Camera (WFCAM) mounted at the United Kingdom Infrared Telescope (UKIRT). For the WFCAM filters, which are very similar to VISTA's (except that WFCAM has $K$ and VISTA has $K_\mathrm{s}$), Hodgkin et al. (\cite{hod09}) have shown that the calibration of the $Y$ band, not included in 2MASS, is possible where the extinction is not too high, i.e. $E(B-V)<1.5$. This is well within the average extinction values towards the Magellanic system (Westerlund et al. \cite{wes97}). However, in star-forming regions the extinction can reach larger values and the calibration may not be reliable. To remedy this situation a calibration based on the observations of standard stars will be produced together with a thorough investigation of star-forming regions from previous data. 

%Zero-points differ among the VIRCAM detectors and will be derived in the Vega system. Zero-points are stable over periods of months until, for example, the sensitivity of the instrument is reduced, by dust on the mirror. 

A high quality global photometric calibration of the VMC survey will
be supported by the homogeneity and accuracy of the 2MASS
catalogue. The best absolute photometry is expected to be accurate to about $1$\%
and, on average, $2$\%, but relative photometry will reach a
much greater accuracy ($\sim$milli-magnitudes for brighter
sources). At periodic intervals and at the end of the survey the
global photometric calibration will be assessed.

\begin{figure}
\resizebox{\hsize}{!}{\includegraphics{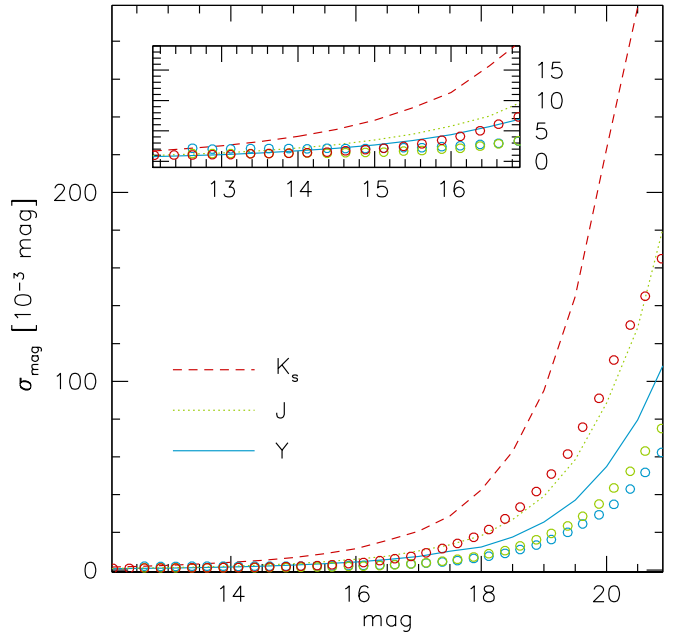}}
\caption{Photometric uncertainties in the VMC data for stacked paw-prints (dashed, dotted and continuous lines in the $K_\mathrm{s}$, $J$ and $Y$ bands respectively) and tiles (red, green and blue circles in the $K_\mathrm{s}$, $J$ and $Y$ bands respectively) in the 8\_8 field. Uncertainties are progressively smaller from $K_\mathrm{s}$ to $Y$ and are systematically smaller in tiles than in stacked paw-prints.}
\label{err}
\end{figure}

Figure \ref{err} shows the behaviour of VISTA photometric
uncertainties in the VMC 8\_8 field where data represent stacked
paw-prints and tiles. Note that the uncertainties reduce by about $\sim$
50\% compared to the individual tiles, and will reduce further for deep
tiles.  In the VMC catalogues several aperture flux magnitudes are
given that sample the curve of growth of all images. The recommended
aperture `aper3', used in this work, corresponds to a core radius of
$1^{\prime\prime}$ ($3$ pixels) that contains $75$\% of the total
stellar flux in a $0.8^{\prime\prime}$ seeing observation. 
%Aperture corrections are calculated assuming that the PSF does not vary across a tile.

\begin{figure}
\resizebox{\hsize}{!}{\includegraphics{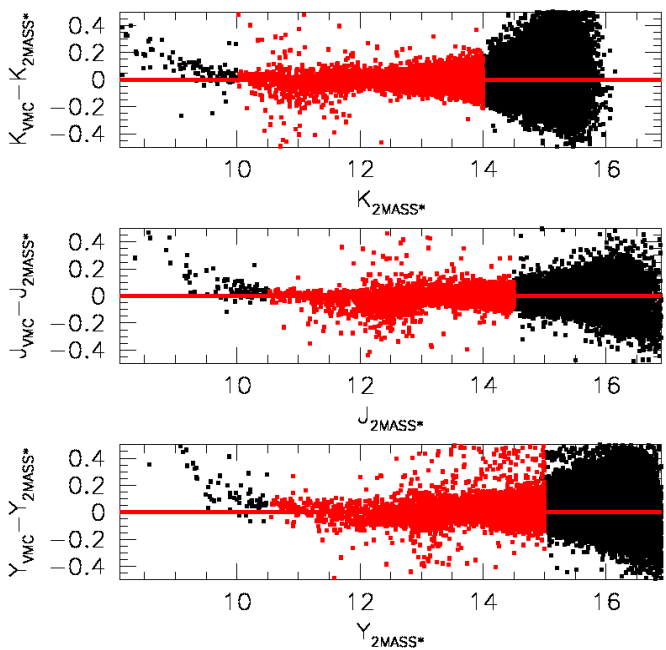}}
\caption{Magnitude difference between VMC and 2MASS$^*$ sources for stars in the 8\_8 field. Horizontal lines were defined within a specific range of magnitudes that is given in the text and is indicated in red in the colour version of the figure.}
\label{phot}
\end{figure}

Figure \ref{phot} shows a comparison between VMC and 2MASS$^*$ magnitudes. By selecting a suitable range of magnitudes the parameters of the comparison (mean, median and sigma) are indicated in Tab. \ref{comp}. Note that by 2MASS$^*$ we do not refer to 2MASS magnitudes, but to the magnitudes obtained using the following colour equations:
\begin{equation}
Y_\mathrm{2MASS^*}=J_{2\mathrm{m}}  +0.550\times(J_{2\mathrm{m}}-H_{2\mathrm{m}}) 
\end{equation}
\begin{equation}
J_\mathrm{2MASS^*} =J_{2\mathrm{m}}  -0.070\times(J_{2\mathrm{m}}-H_{2\mathrm{m}})
\end{equation}
\begin{equation}
K_\mathrm{2MASS^*} =K_{2\mathrm{m}}  +0.020\times(J_{2\mathrm{m}}-K_{2\mathrm{m}}) 
\end{equation}
where $J_{2\mathrm{m}}$, $H_{2\mathrm{m}}$ $K_{2\mathrm{m}}$ are 2MASS magnitudes. These are the formulas used to calibrate the VISTA photometry\footnote{http://casu.ast.cam.ac.uk/surveys-projects/vista/technical/vista-sensitivity}. They include the correct colour term but do not include the small zero-point shifts that are necessary to bring the observations into a Vega magnitude system.

\begin{table}
\caption{VMC-2MASS$^*$ comparison.}
\label{comp}
\[
\begin{array}{ccccc}
\hline
\noalign{\smallskip}
\mathrm{Band} & \mathrm{Range} & \mathrm{Mean} & \mathrm{Median} & \mathrm{Sigma} \\ 
 & \mathrm{(mag)} & \mathrm{(mag)} & \mathrm{(mag)} & \mathrm{(mag)} \\
\hline
\noalign{\smallskip}
 Y & 10.5-15.0 & +0.018 & +0.012 & 0.110 \\
 J  & 10.5-14.5 & -0.005 & -0.003 & 0.092 \\
 K_\mathrm{s} & 10.0-14.0 & -0.003 & -0.003 & 0.069 \\
\noalign{\smallskip}
\hline
\end{array}
\]
\end{table}

The morphological classification is encoded in a flag with values: $-1$ for stellar, $+1$ for non-stellar, $0$ for noise and $-2$ for borderline stellar sources. This classification indicates the most probable source morphology and is affected by crowding and sensitivity. For details about the photometric apertures, the source classification and other catalogue parameters the reader is referred to the CASU web pages.

\subsection{Image quality}
\label{quality}

The quality of the VMC images is evaluated at three different steps --
during the observations, the data reduction, and the archiving
process. The VMC team performs additional quality control checks that
interleave with each of these three steps. The first identifies
obvious causes for re-observation, such as observations that exceeded the
required constraints or that failed to be completed
because of technical reasons. These observations are usually repeated
almost immediately (Tabs. \ref{otab}, \ref{ktab}). All images are
processed by CASU and archived at the VSA, regardless of whether their
observing constraints are met.

All CASU-reduced VMC survey images for individual detectors have been inspected visually for quality control. The purpose of this inspection is to recognise artefacts and residuals from the reduction process but also to identify features that are intrinsic to the observations. The results of the quality inspection are as follows:
\vspace{-0.2cm}
\begin{itemize}
\item The upper $1/3$ of detector \#$16$ is effectively noisy in the bluer bands, this also causes a calibration problem for that region: so the VMC $Y$ band observations suffer more than the $J$ and $K_\mathrm{s}$ ones. This problem causes an increasing background level and influences the detection capabilities. 
\item The observations obtained before $20$ November $2009$ suffer from intermittent problems in detector \#$6$ channel \#$14$ (each detector has $16$ channels, processed by different analog-to-digital convertors) that required replacement of a video board in the controller. These stripes cover an area comparable to that of a bright foreground star. 
\item A special sky frame needs to be used for reducing images obtained during the night of $19$ November $2009$, due to investigations on the component responsible for the previous problem. This step is implemented in the pipeline reduction from version $0.9$ onwards.
\item A low quality region at the bottom (`$-$Y') of detector \#4 creates a horizontal pattern that does not cancel out with stacking images obtained from the exposure and jittering sequence. This problem will likely not affect subsequent reductions of the data.
\item Overall the reduced images show a smooth gradient most noticeable in the $K_\mathrm{s}$ band (possibly caused by the baffling system of VIRCAM as a result of thermal radiation), but does not present a problem for the source extraction.
\end{itemize}
\vspace{-0.2cm}
None of these problems require re-observation of the tiles that have been obtained for the VMC survey.\\

The quality of the images was further inspected by comparing the FWHM of the extracted sources with the expected seeing requirements. As expected 
the FWHM varies among the VIRCAM detectors. For example, the average FWHM in the corners of detector \#$1$ is larger than in other detectors. This implies that some detectors will have a FWHM that exceeds the seeing requirement by more than $10$\%. The average seeing among the detectors and the average seeing among the six paw-prints of a tile is, however, always within the required limit, except in a few cases, see Tabs. \ref{otab} and \ref{ktab}. 

Tables \ref{otab} and \ref{ktab} also show a few cases where an insufficient number of paw-prints or jitters are present and tiles are not fully sampled. By discarding these problematic observations among those that remain only very few need to be re-obtained. Table \ref{avgpar} shows the average seeing, ellipticity, zero-point and limiting magnitude from all VMC images. These parameters have been calculated excluding problematic observations and represent average values regardless of their execution as part of a concatenation, group or monitoring sequence, as well as from the specific requirements of the different VMC fields with respect to crowding.\\

\begin{table}
\caption{Average VMC parameters from all single tile images.}
\label{avgpar}
\[
\begin{array}{ccccc}
\hline
\noalign{\smallskip}
\mathrm{Band} & \mathrm{FWHM} & \mathrm{Ellipticity} & \mathrm{Zero-point} & 5\sigma\, \mathrm{Mag.\,Limit} \\ 
 & \mathrm{(arcsec)} & & \mathrm{(mag)} & \mathrm{(mag)} \\
\hline
\noalign{\smallskip}
 Y & 1.03\pm0.13 & 0.065\pm0.011 & 23.520\pm0.070 & 21.111\pm0.395 \\
 J  & 1.00\pm0.10 & 0.064\pm0.011 & 23.702\pm0.206 & 20.527\pm0.382 \\
 K_\mathrm{s} & 0.93\pm0.08 & 0.051\pm0.009 & 22.978\pm0.245 & 19.220\pm0.340 \\
\noalign{\smallskip}
\hline
\end{array}
\]
\end{table}

\section{Data archive}
\label{vsa}

The data reduced by the VDFS pipeline at CASU are ingested into the VSA\footnote{http://horus.roe.ac.uk/vsa/login.html} at the Wide Field Astronomy Unit (WFAU) in Edinburgh which is similar to the WFCAM Science Archive (Hambly et al. \cite{ham08}).
At present, these are data reduced with v$1.0$ of the CASU pipeline and include all VMC data observed until end of May $2010$ (Tabs. \ref{otab} and \ref{ktab}). At VSA the data are curated to produce standardised data products. The software that runs at WFAU and populates the VSA is the same that runs at CASU and this guarantees that the data are processed homogeneously throughout the entire processing chain.

The most important processes, available at present, for the VMC survey are: individual passband frame association and source association to provide multi-colour, multi-epoch source lists; cross-association with external catalogues (list-driven matched photometry);  deeper stacking in specified fields; quality control procedures. 

There are three main types of VSA tables that are important for the
VMC survey. These are the {\it vmcDetection} table, the {\it
vmcSource} table and the {\it vmcSynoptic} table(s). The vmcDetection
table contains the catalogues corresponding to individual
observations. At the moment there is one catalogue per paw-print,
regardless of band and tile of origin. The vmcSource table contains
the list of sources obtained from deep stack images and each source is
matched in the three VMC bands. In practise, each row of the vmcSource
table will contain $Y$, $J$ and $K_\mathrm{s}$ magnitudes for a
source. At present, because VSA is organised by paw-prints, the same
source may appear two or more times in the vmcSource table depending
on its location with respect to the overlap among the six paw-prints
forming a tile. The synoptic tables contain the colour information and
the multi-epoch information for individual observations (a single
OB). More details about the synoptic tables are given by Cross et
al. (\cite{cro09}).  The position and magnitude for each source in any
given table refers to the astrometrically and photometrically calibrated
measurements using the parameters specified in the image
headers. These parameters are discussed in Sect. \ref{data}. In
addition, there are several quality flags that are specifically
introduced at the VSA level. These flags identify problems occurring
during the ingestion of the data into the archive, incompleteness in
the set of data (for example missing exposures in a paw-print
sequence), problems related to the pairing of data, etc.

\begin{figure}
\resizebox{\hsize}{!}{\includegraphics{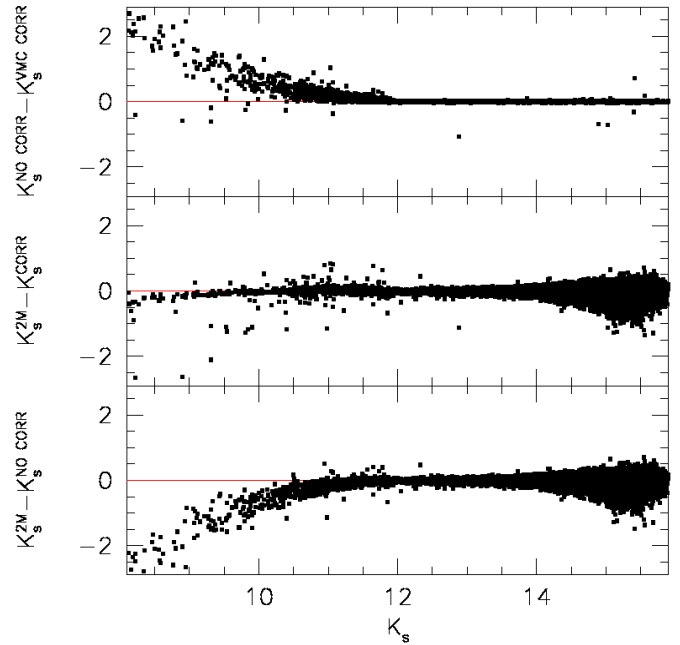}}
\caption{Magnitude difference for stars in the 8\_8 field in common between VMC and 2MASS before (bottom) and after (middle) correcting the magnitude of VMC stars approaching the saturation limit, and (top) the correction itself developed by Irwin (\cite{irw09}).}
\label{sat}
\end{figure}

The magnitudes of the brightest stars are corrected for saturation
effects (Sect. \ref{saturation}). Figure \ref{sat} shows the
$K_\mathrm{s}$ magnitude difference for stars in the 8\_8 field
compared to 2MASS before and after saturation correction. Saturation
effects are present for $K_\mathrm{s}< 12$ mag, with the magnitudes
of brighter stars recovered to at least $K_\mathrm{s}=10$ mag after
correction.

The VSA is queried using Structured Query Language (SQL) and a point-and-click web form for browsing. This is a dual (sophisticated and simple) end-user interface for the data. A key feature to note is the design with multi-waveband catalogue data that allows the user to track back to the individual source images and merged-source tables, and present the user with a generally applicable, science-ready dataset. The VSA has a high-speed query interface, links to analysis tools such as TopCat, and advanced new VO services such as MySpace. 
The VSA supports a different range of queries and the most common for the VMC survey are: (i) querying the archive to check which data have been ingested, (ii) querying the vmcSource table to extract magnitudes from the deep stacks, (iii) querying the synoptic tables to extract light-curves and statistics on the levels of variability and (iv) querying the VSA using an input list of sources and searching for their VMC counterpart. For each source in (i) it is possible to inspect postage stamp images in each wave band.

Before creating the deep stack images, the quality control results discussed in Sect. \ref{quality} need to be taken into account. In particular, all images except those with a large seeing and/or ellipticity listed in Tabs. \ref{otab} and \ref{ktab} are included in the deep stacks. This means that tile images with a reduced number of jitters or paw-prints are included into the deep stacks if their observing conditions are met. The latter will still be available as individual epochs, and will be linked to the other observations in the synoptic tables, because they may contain useful information for variable stars or for source confirmation. 

In the future it may be possible to automatically join the under-exposed areas at the two ends of each tile with those of the adjacent tiles prior to source extraction. The VSA also contains external catalogues, like 2MASS, that can be linked with the VMC data via an SQL query. Catalogues that are specifically important for the Magellanic system, e.g. the MCPS and the SAGE catalogues are also being ingested into the VSA. VMC is intrinsically a multi-wavelength project and a large fraction of its science will come from the linking of VISTA data with those from other surveys; the VSA is designed to enable such links. 

\section{Analysis and Results}
\label{results}

\begin{figure*}
\resizebox{\hsize}{!}{\includegraphics{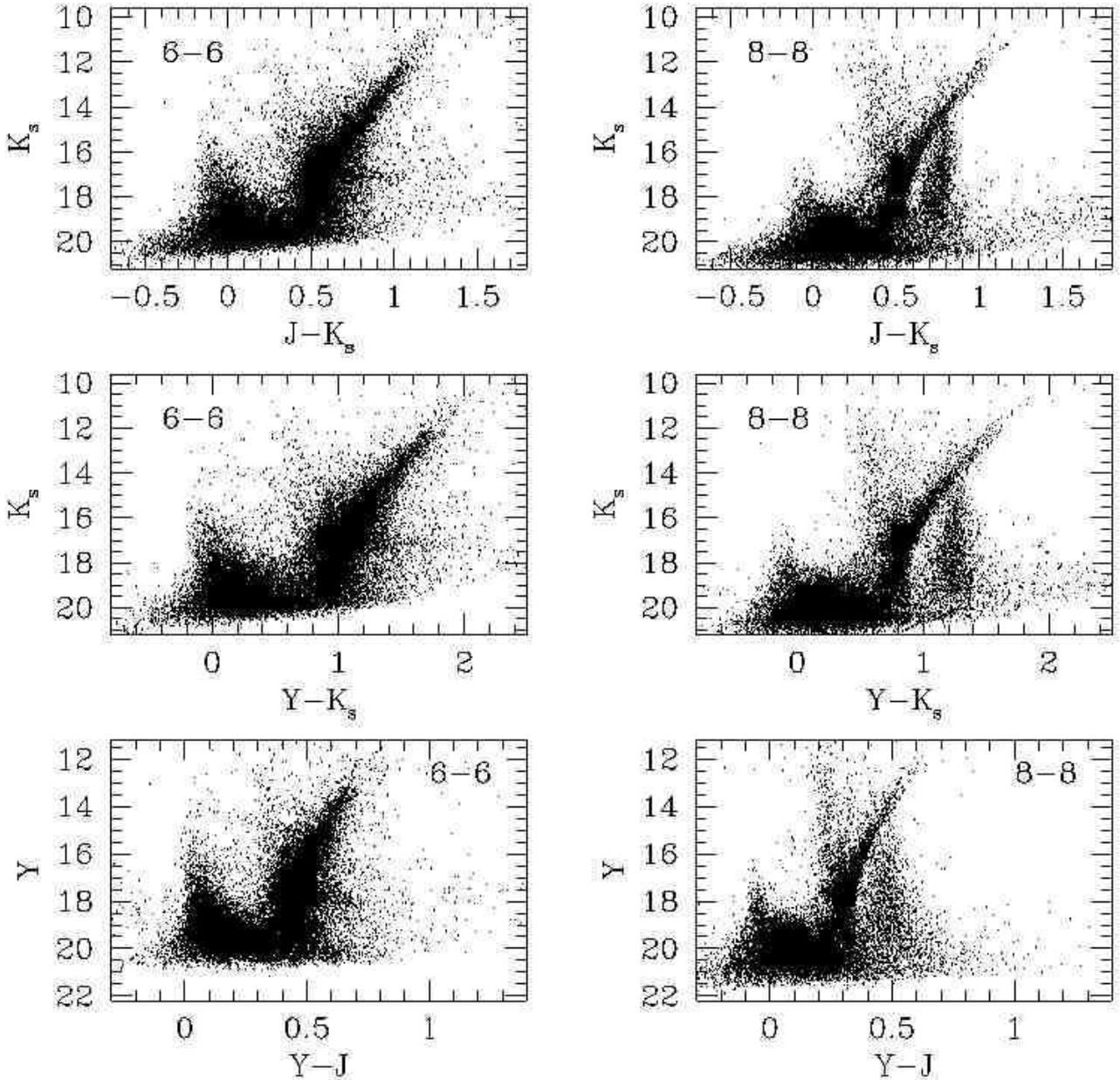}}
\caption{Colour-magnitude diagrams of VMC sources in part of the 6\_6 (30 Dor) and 8\_8 (Gaia SEP) LMC fields.}
\label{fig6}
\end{figure*}

Figure \ref{fig6} shows the CMDs of the VMC data in the 6\_6 (30 Dor) and 8\_8 (Gaia) LMC fields. These data were extracted from the VSA. The magnitudes and colours of each source correspond to a single detector, i.e. if the same source was detected in another detector it is not included in the CMDs. This is because at this stage the archiving process is organised by paw-prints. When tiles, resulting from the combination of six paw-prints, become available at VSA then the exposure time per source will be at least doubled. At present the exposure times per band for the sources shown in the CMDs correspond to $1200$ sec in $Y$, $1400$ sec in $J$ and to $\sim 4000$ sec in $K_\mathrm{s}$.

The distribution of stars in the CMDs shows clearly the different
stellar populations characterising these LMC fields. The blue-most
conic structure bending to red colours at bright magnitudes is formed
by main-sequence (MS) stars of increasing mass with increasing
brightness. The MS joins, via the sub-giant branch, the RGB beginning
at $\sim$2 mag below the red clump, the approximately circular region
described by the highest concentration of stars. The structure of the
red clump depends on stellar parameters (age and metallicity) but also
on interstellar extinction. Extinction causes the clump to elongate to
red colours, as seen in the CMDs for the 30 Dor field where
extinction is higher than in the Gaia field. The RGB continues beyond
the red clump at brighter magnitudes describing a narrow structure
bending to red colours. The abrupt change in source density at the tip
of the RGB marks the transition to brighter AGB stars. The broad
vertical distribution of stars below the RGB is populated by MW
stars. In the CMDs of the Gaia field these are easily distinguished
from LMC stars. Cepheid and supergiant stars occupy the region of the
diagram to the bright and blue side of the RGB while RR Lyrae stars
are somewhat fainter than the red clump and lie more or less parallel
to the sub-giant branch.

\begin{figure}
\resizebox{\hsize}{!}{\includegraphics{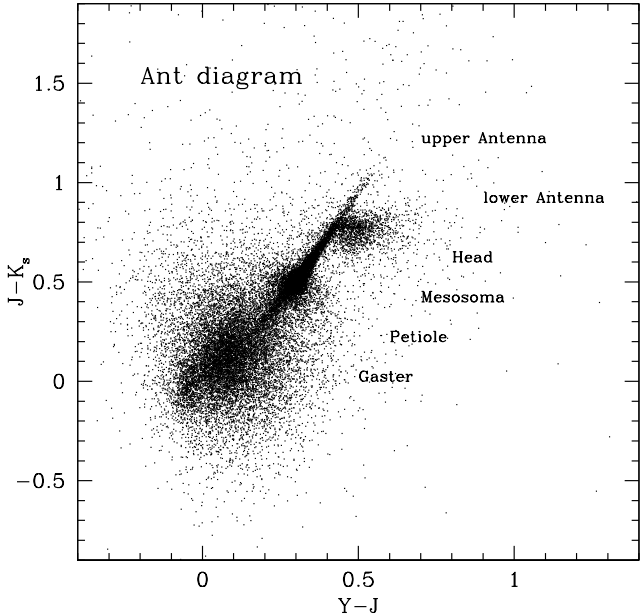}}
\caption{Ant diagram: the colour-colour diagram of VMC sources in part of the 8\_8 LMC field.}
\label{ant}
\end{figure}

Figure \ref{ant} shows the colour-colour diagram of the VMC data in the 8\_8 field. The data shown here are the same as in the CMDs (Fig. \ref{fig6}) described above. The distribution of sources in the colour-colour diagram resembles the body of an ant. Following this analogy:
\vspace{-0.1cm}
\begin{itemize}
\item The lower part of the ant body ({\it gaster}) at $-0.1<(Y-J)<0.2$, $-0.3<(J-K_\mathrm{s})<0.2$, corresponds to the location of MS stars in the LMC, with the youngest stars being at the bluest extremity. The width of this feature is mainly caused by photometric uncertainties. Its extension to the red is also set by the limited depth of the VMC data, since the faint MS should continue to even redder colours.

\item The middle part of the ant body ({\it mesosoma}) at $0.2<(Y-J)<0.4$, $0.3<(J-K_\mathrm{s})<0.65$, corresponds to the main locus of helium-burning giants in the LMC. The bulk of them are in the red clump (see Fig. \ref{fig6}), but also brighter helium burning giants, and stars in the faint extension of the red clump, fall in this same blob.

\item The {\it petiole} is a small thin extension of the mesosoma at its red side, at $0.2<(Y-J)<0.3$, $0.2<(J-K_\mathrm{s})<0.3$, and is mainly caused by bright stars in the MW foreground -- more specifically by the intermediate-age and old turn-offs of MW disk populations (at  $K_\mathrm{s}<15$).
\end{itemize}

In addition to this well-defined petiole, gaster and mesosoma are also connected by the relatively less populated LMC sub-giants, lowest-luminosity RGB stars, and horizontal branch stars in the LMC.

\vspace{-0.2cm}
\begin{itemize}
\item The upper part of the ant body ({\it head}) is a more complex feature. Its main blob at $0.4<(Y-J)<0.6$, $0.6<(J-K_\mathrm{s})<0.8$ is defined by low-mass stars in the MW foreground, especially those with masses $<0.5$ M$_\odot$ which clump at the same near-IR colours ($J-K_\mathrm{s}\sim 0.7$; see Nikolaev \& Weinberg \cite{nik00}, Marigo et al. \cite{mar03}). The same structure forms a marked vertical feature in the CMDs.

\item Two {\it antennae} depart from this head, the upper one at $0.5<(Y-J)<0.6$, $0.8<(J-K_\mathrm{s})<1$ being formed by the more luminous RGB stars in the LMC, close to their tip of the RGB, extending up to $(J-K_\mathrm{s})=1$ mag. This upper antenna finishes abruptly because the tip has been reached. The lower antenna at $(J-K_\mathrm{s})\sim 0.8$, $(Y-J)>0.6$, is more fuzzy, and corresponds to the $Y-J$ red-ward extension of low-mass stars in the MW foreground. This red ward extension is partially caused by photometric uncertainties and by the particular colour-colour relation followed by the coolest M dwarfs.
\end{itemize}

\subsection{Completeness}

Some key science goals of the VMC survey require accurate estimates of the completeness of the stellar photometry as a function of location across the Magellanic system, and position in the CMDs. This is estimated via the usual procedure of adding artificial stars of known magnitudes and positions to the images, then looking for them in the derived photometric catalogues. 

\begin{figure}
\resizebox{\hsize}{!}{\includegraphics{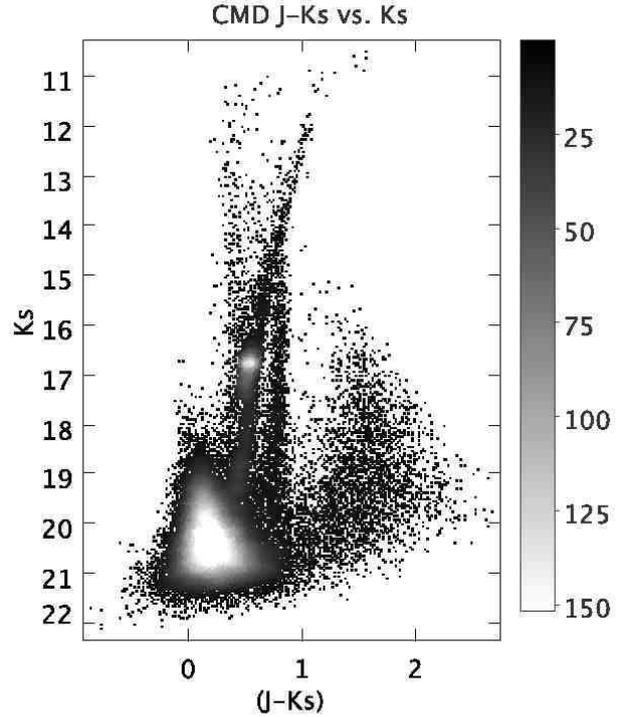}}
\caption{CMD for a region of $4000\times 4000$ pixels extracted from the 8\_8 LMC deep tile. Objects with $J-K_\mathrm{s}>1$ and fainter than $K_\mathrm{s}=16$ mag are background galaxies.}
\label{cmd1jk}
\end{figure}

For this work, the paw-print images, available for the 8\_8 LMC field, were combined to produce a tile image using the SWARP tool (Bertin et al. \cite{ber02}). A region of the tile image with a size of $4000\times 4000$ pixels was selected. Then, PSF photometry using the DAOPHOT and ADDSTARS packages in IRAF was performed. Figure \ref{cmd1jk} shows an example of the photometry on the (preliminary) stacked image. 

The artificial stars are positioned at the same random position on the $YJK_\mathrm{s}$ images. Their mutual distances are never smaller than $30$ pixels, so that the process of adding stars does not increase the typical crowding of the image. Later on, artificial stars spanning small bins of colour and magnitude are grouped together to provide estimates of number ratio between added and recovered stars -- i.e. the completeness -- as a function of position in the CMDs.

\begin{figure}
\resizebox{\hsize}{!}{\includegraphics{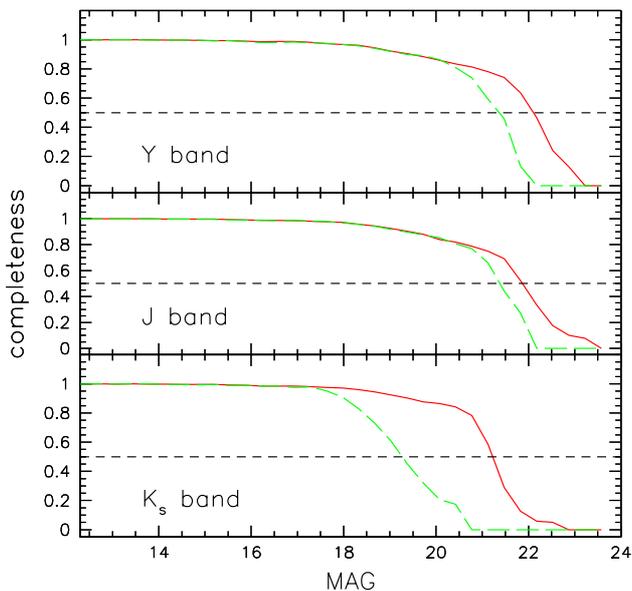}}
\caption{Completeness results for the 8\_8 LMC tile with a single epoch (dashed line) and with deep stacked images (continuous line).}
\label{completeness}
\end{figure}

Typical results of this process are illustrated in the
Fig. \ref{completeness}, which shows the completeness as a function of
magnitude, for both cases of single epoch and deep stacked images. For
this relatively low density tile, the figure shows that the $\sim$50\%
completeness limit goes as deep as $22.20$, $21.90$, and $21.40$ mag
in the $Y$, $J$ and $K_\mathrm{s}$ bands, respectively. These results
are in good agreement with the expectations derived from simulated VMC
images (Kerber et al. \cite{ker09}).

In order to derive the spatially-resolved SFH of the Magellanic
Clouds, we are performing a more extensive series of artificial star
tests in the available images. The results will be discussed in detail
in a forthcoming paper of this series.

\subsection{RR Lyrae stars and Cepheids}
\label{variables}

Radially pulsating stars obey a period-mean density relation that forms the basis of their use as standard candles to measure distances to the host systems. RR Lyrae stars in particular, obey a period-luminosity-metallicity (PLZ) relation in the $K$ band which is weakly affected by evolutionary effects, spread in stellar mass within the instability strip, and uncertainties in reddening corrections (see Longmore et al. \cite{lon86}, and Sollima et al. \cite{sol06} for updates). Similarly, the Cepheid  PL relation in the $K$ band is much narrower than the corresponding optical relations, and less affected by systematic uncertainties in reddening and metal content (Caputo, Marconi \& Musella \cite{cap00}).

The $K$-band PLZ relation of RR Lyrae stars has  already been used by a number of authors to derive the distance to LMC clusters and field regions (Dall'Ora et al. \cite{dal04}, Szewczyk et al. \cite{sze08}, Borissova et al. \cite{bor09}). However, these studies only cover a few tiny LMC portions mainly located in or close to the LMC bar, and just one LMC cluster, the Reticulum  (Dall'Ora et al. \cite{dal04}). In the context of the VMC project, the $K_\mathrm{s}$ photometry is taken in time series mode in order to obtain mean $K_\mathrm{s}$ magnitudes for RR Lyrae stars and Cepheids over the whole MCs, and use their PL relations to measure distances and construct a 3D map of the entire system. 

Over the last decades, large microlensing surveys such as MACHO, EROS
and OGLE have been conducted in the optical domain to search for
planetary-to-stellar-mass baryonic dark matter in the MW. As a
by-product these surveys have produced tens of millions of light
curves of individual background stars. These surveys provide an
unprecedented opportunity for systematic studies of whole classes of
variable stars and their host galaxies. They cover different fractions
of the MC area. Specifically, the MACHO survey (see Alcock et
al. \cite{alc04} and references therein) cover a region mainly
encompassing the LMC and SMC bars. The OGLE surveys (Soszy\'{n}ski et
al. \cite{sos08,sos09} and references therein) of which stage IV is in
progress, cover a progressively larger area extending further outside
the bar of each Cloud. EROS-2 (Tisserand et al. \cite{tis07}) is the
most extended of these surveys and, at present, is the one covering
the largest fraction of the VMC survey area.

We are using the identification, the period and the epoch of maximum
light of RR Lyrae and Cepheids, identified by the microlensing
surveys, to fold the $K_\mathrm{s}$-band light curves produced by VMC
and derive average $K_\mathrm{s}$ magnitudes for these variables. In
this paper we present results of a preliminary analysis based on the
combination of VMC, EROS-2 and OGLE III data for Cepheids and RR Lyrae
stars in two LMC fields, namely, the Gaia and 30 Dor fields.

Given its location on the periphery of the LMC, the Gaia field only
overlaps with EROS-2.  Coordinates, periods and optical\footnote {The
EROS $blue$ channel (420-720 nm), overlapping the $V$ and $R$ standard
bands, and $red$ channel (620-920 nm), roughly matching the mean
wavelength of the Cousins $I$ band (Tisserand et al. \cite{tis07})}
light-curve dataset of RR Lyrae and Cepheids in the Gaia field are 
taken from the EROS-2 catalogue and cross-matched to the VMC 
data. The 30 Dor field is covered by both EROS-2 and OGLE III, as well as by
MACHO, but for the present analysis we only employ periods and optical
(Johnson-Cousins $V, I$ bands) light curves from OGLE III
(Sosz\'{n}ski et al. \cite{sos08,sos09}).

RR Lyrae stars and Cepheids in the Gaia field were extracted from the
EROS-2 catalogue and the ($B-R, R$) CMD using the following limits:
$18.46<B<20.03$ mag, $0.05<(B-R)<0.58$ mag for the RR Lyrae stars, and
$13.39<B<17.82$ and $0.89<P<15.85$ days for the Cepheids. This
selection returned a list of $16337$ RR Lyrae stars and $5800$
Cepheids over the whole field of the LMC covered by EROS-2. RR Lyrae
candidates and Cepheids within the Gaia field were then extracted
according to their coordinates by considering only objects with $
87.8464 < \alpha < 91.8464$ deg and $ -67.3413 < \delta < -65.3413$
deg, giving $235$ RR Lyrae and $47$ Cepheid candidates. 
The RR Lyrae candidates were then further selected by considering only
objects with `proper' periodicities from EROS-2. This restricted the
sample to $218$ sources inside the Gaia field. A preliminary
cross-match between this catalogue and the VSA\footnote{Database
release VMC20100607} was then made using the $CrossID$ query (pairing
radius $1^{\prime\prime}$), yielding a final catalogue of $117$ RR Lyrae
candidates and $21$ Cepheids in common.

Astrometric differences between the EROS-2 sources and VMC
counterparts is $<1^{\prime\prime}$ for $98.4$\% of the sources (with
differences for the remainder in excess of $5^{\prime\prime}$).
Periods for these stars were checked by analysing their
$B_\mathrm{EROS}$ light-curves with Graphical Analyser of Time Series
(GRATIS) custom software developed at the Bologna Observatory by
P. Montegriffo (Clementini et al. \cite{cle00}) confirming the EROS-2
periodicities for the majority. Similar methods were employed to
select and cross-match RR Lyrae stars and Cepheids in the 30 Dor field
from the OGLE III catalogue, and to extract their $K_\mathrm{s}$ band
time series data from the VMC observations.

\begin{figure*}
\resizebox{22cm}{10cm}{\includegraphics{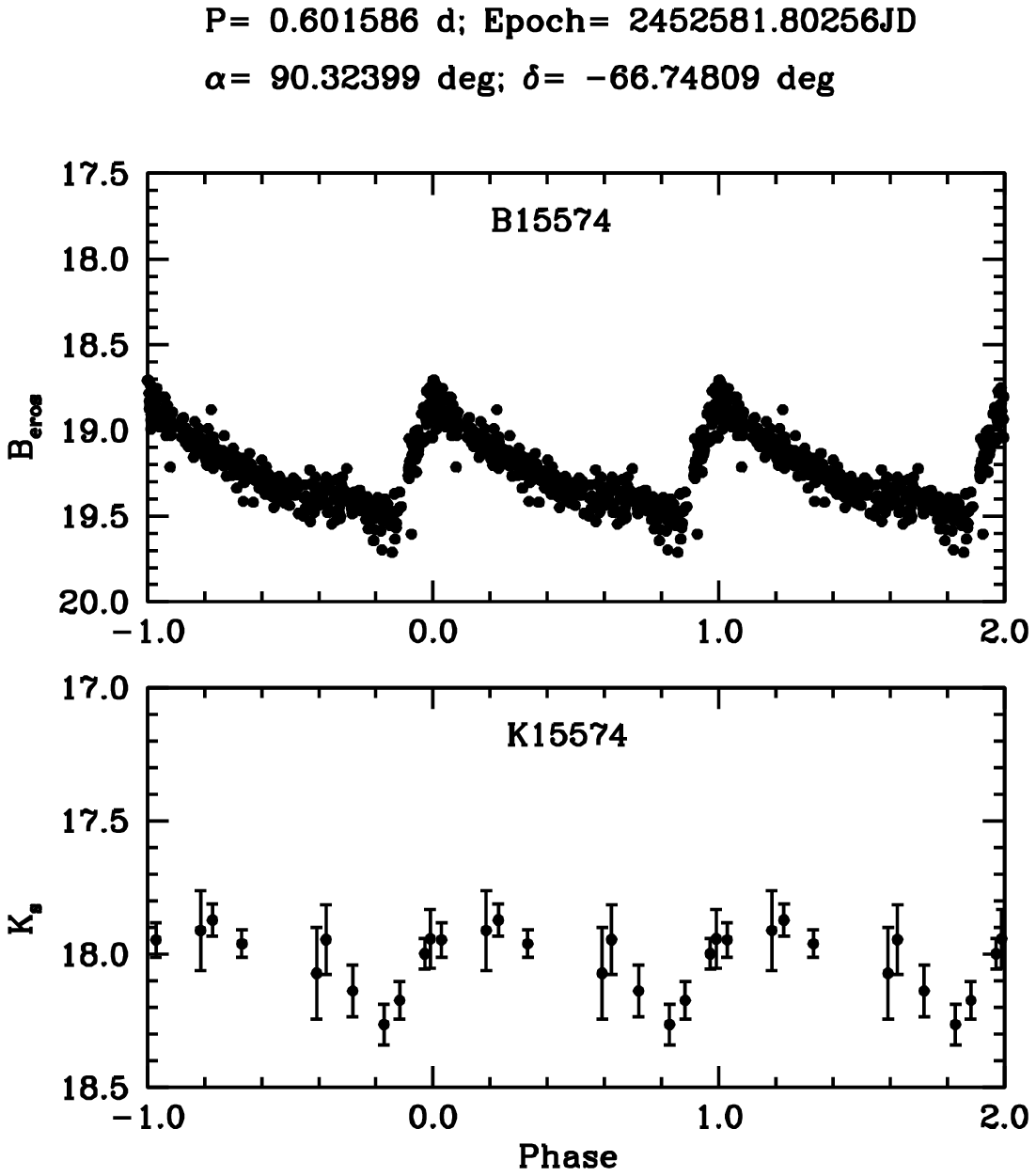}
\hspace{-9cm}
\includegraphics{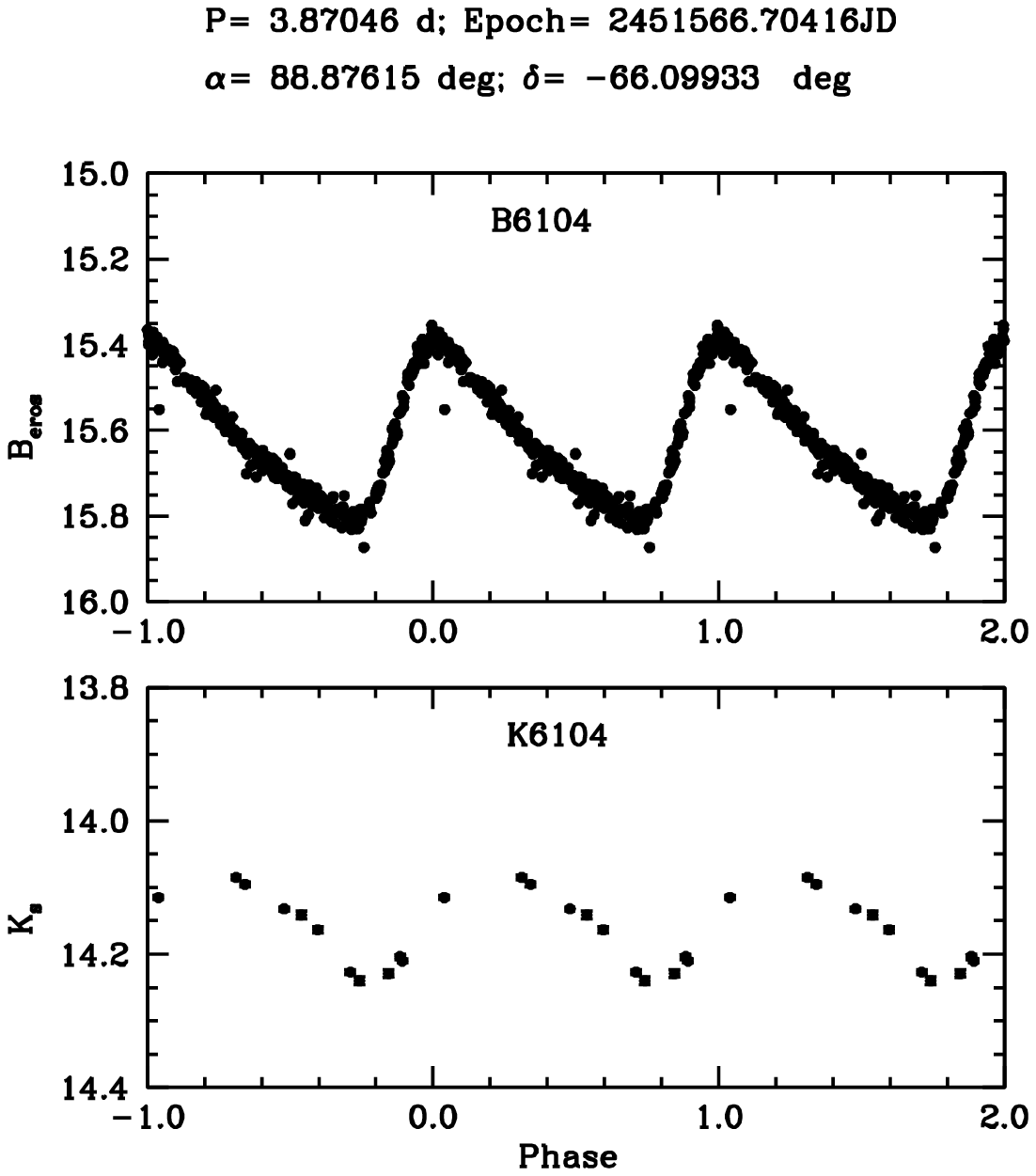}}
\resizebox{22cm}{10cm}{\includegraphics{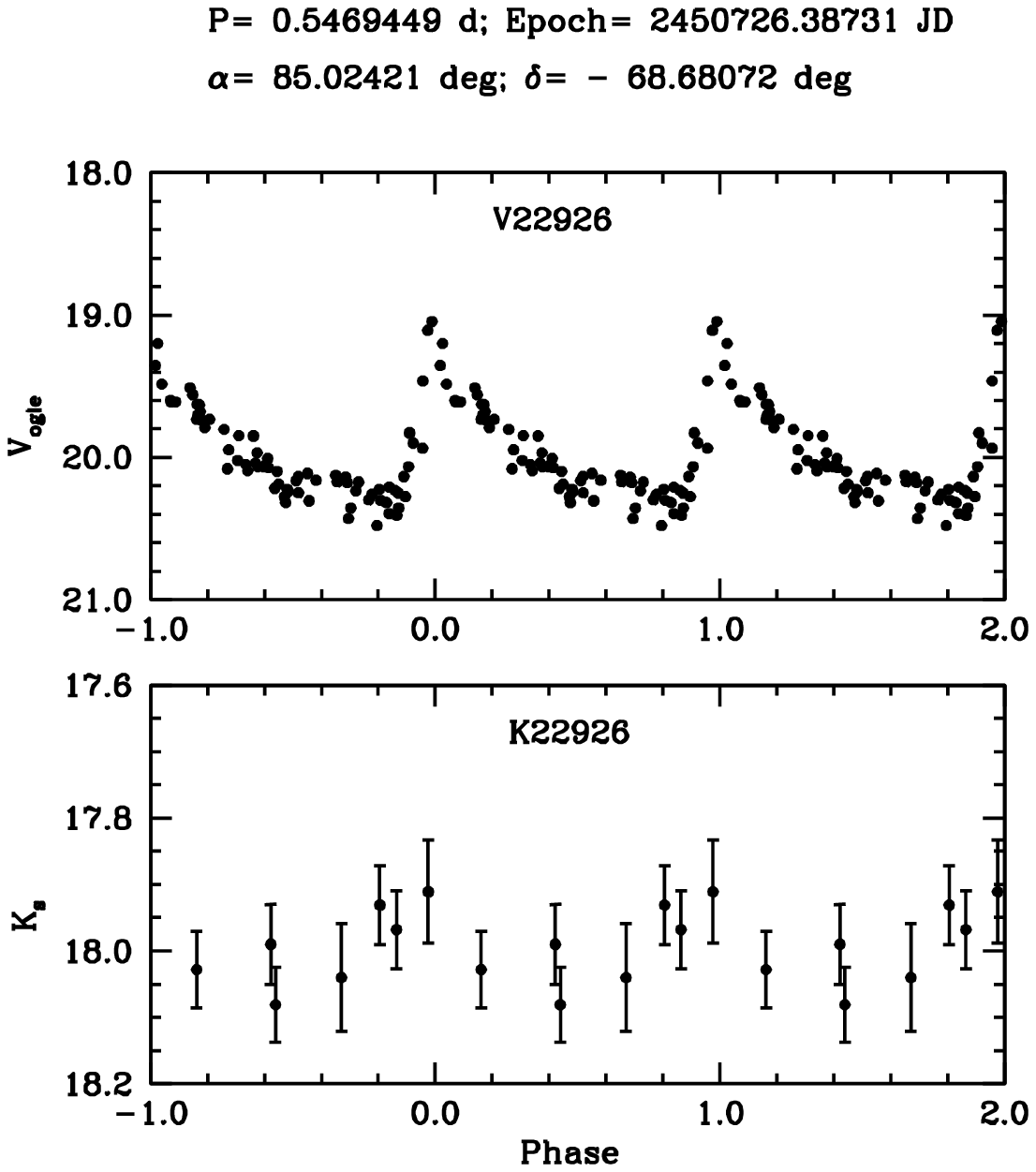}
\hspace{-9cm}
\includegraphics{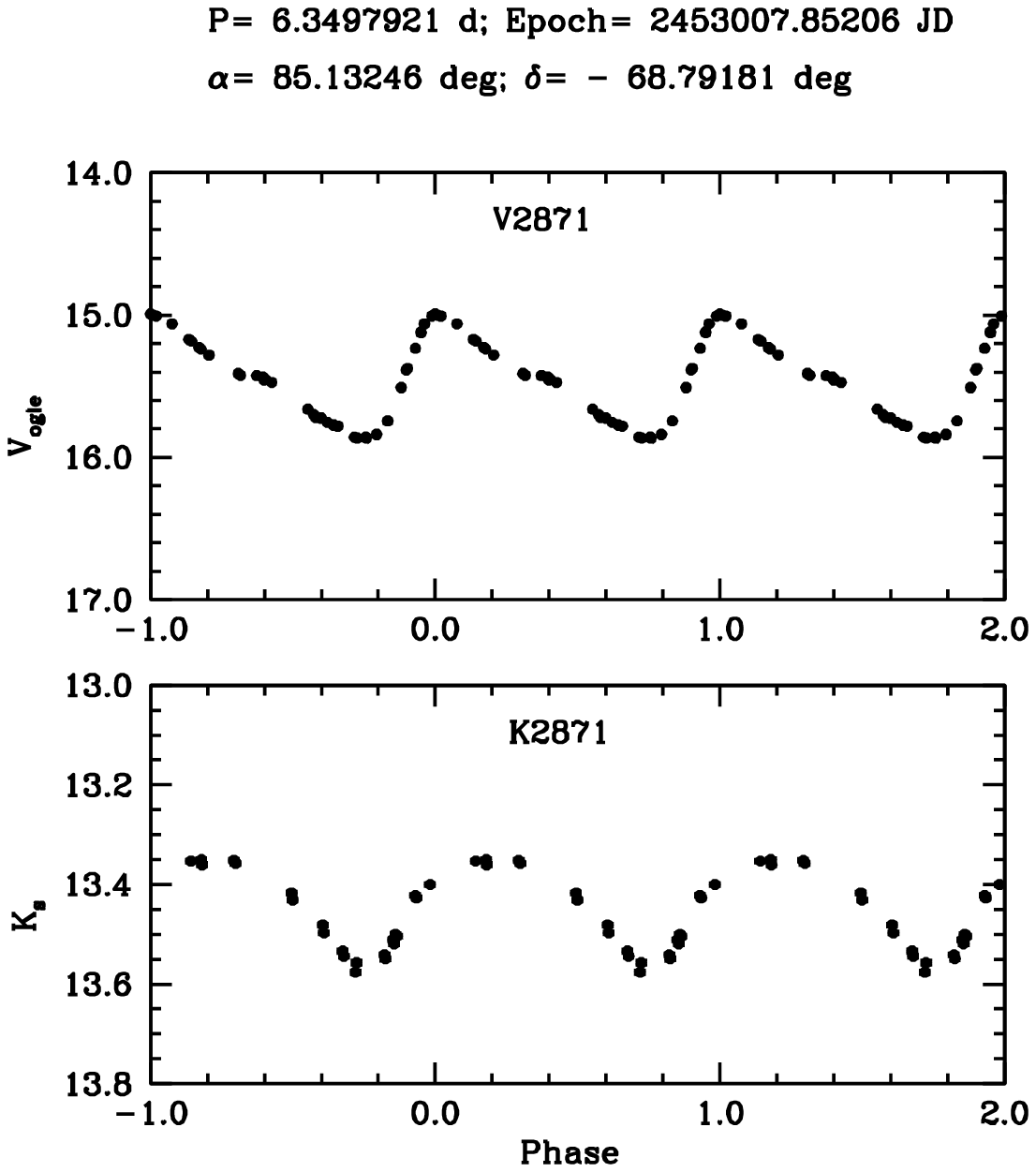}}
\caption{$B_\mathrm{EROS}$ and $K_\mathrm{s}$ VMC light-curves of the fundamental mode RR Lyrae  star \#15574 and for Cepheid \# 6104 in the Gaia SEP field (top) as well as $V_\mathrm{OGLE~III}$ and $K_\mathrm{s}$ VMC light-curves of the fundamental mode RR Lyrae  star \#22926 and for Cepheid \# 2871 in the 30 Dor field (bottom). Data are folded according to the period and epoch of maximum light derived from the EROS-2 data.}
\label{fig:curve}
\end{figure*}

Figure \ref{fig:curve} shows $B_\mathrm{EROS}$ and $K_\mathrm{s}$ VMC
light curves of a fundamental mode RR Lyrae star (star \#15574, with
$P=0.601586$ days) and a Cepheid (star \#6104, with $P=3.87046$ days)
in the Gaia field. Also shown are $V_\mathrm{OGLE~III}$ and
$K_\mathrm{s}$ VMC light-curves of a fundamental mode RR Lyrae star
(star \#22926, with $P=0.5469449$ days) and a Cepheid (star \#2871,
with $P=6.3497921$ days) in the 30 Dor field\footnote{The variables in
the Gaia field have EROS-2 identifications of lm0382m15574 and
lm0507l6104 for the RR Lyrae and Cepheid, respectively.  Similarly,
the variables in the 30 Dor field have OGLE III identifications of
OGLE-LMC-RRLYR-22926 and OGLE-LMC-CEP-2871, for the RR Lyrae and
Cepheid, respectively.}.

In practise, in the VMC paw-print the same source will (on average)
appear in at least one more detector within the same tile sequence;
the $K_\mathrm{s}$ magnitudes and uncertainties shown in
Fig. \ref{fig:curve} for RR Lyrae stars are the weighted averages of
all available observations, so for each night we have one point. For
the Cepheids all the $K_\mathrm{s}$ data available in the VSA release
(VMC20100607) are shown in the figure, i.e., two points for each night,
in order to show the excellent raw data for these stars.

Error bars of the individual $K_\mathrm{s}$ measurements are shown in
the figures. The uncertainties for the Cepheids in both fields are the
same size at the data-points in the figures, while they are larger for
the 30 Dor RR Lyrae star due to the crowding in this part of the
LMC. Nevertheless, the typical shape of the fundamental mode of the
RR Lyrae stars is easily recognised in Fig. \ref{fig:curve}. The $K_\mathrm{s}$
light-curves are very well sampled for both Cepheids and RR Lyrae
stars confirming the soundness of our observing strategy, and allowing
us to derive accurate $\langle K_\mathrm{s} \rangle$ magnitudes
without using template light curves. The $\langle K_\mathrm{s}
\rangle$ magnitudes, computed as the average of the individual
$K_\mathrm{s}$ measures correspond to $18.02 \pm 0.12 $ mag, $17.99
\pm 0.06 $ mag for the Gaia and the 30 Dor RR Lyrae stars,
respectively, and to $14.17 \pm 0.06 $ mag, $13.46 \pm 0.08 $ mag for
the Gaia and the 30 Dor Cepheids, respectively where the quoted 
uncertainties are the standard deviations of the averages.

The $\langle K_\mathrm{s} \rangle$ magnitudes from the VMC time series
will be used along with EROS-2 and OGLE III light-curves to construct
${\rm PL_{\it K}Z}$ relations for the RR Lyrae stars and PL, PL-colour
and Wesenheit relations for Cepheids, to then investigate the
geometrical distribution of variables in these LMC fields and for the
MC system in general.

\subsection{Planetary nebulae}

There are approximately $700$ objects catalogued as PNe in the LMC (Leisy et al. \cite{lei97}; Reid \& Parker \cite{rei06a}, \cite{rei06b}) and $140$ in the SMC (Jacoby, private communication; Sanduleak \& MacConnell \& Philip \cite{san78}; Jacoby \cite{jac80}; Jacoby \& De Marco \cite{jac02}). Magellanic PNe are best known for their fundamental role in the development of the extragalactic standard candle [O\,{\footnotesize III\normalsize}] $\lambda 5007$ Planetary Nebula Luminosity Function (PNLF; Henize \& Westerlund \cite{hen63}; Jacoby \cite{jac80}, \cite{jac89}). Distances can be measured from the near-universal bright end cut-off across all galaxy types, but it remains difficult to explain how old stellar populations that lack recent star-formation episodes could produce progenitors massive enough to power the high central star luminosities populating the bright end (Jacoby \cite{jac97}, Marigo et al. \cite{mar04}, Ciardullo \cite{cia10}). 

Some success in reproducing the observed PNLF has been achieved by incorporating the latest hydrodynamic, time-dependent models (Sch\"onberner et al. \cite{sch07}, M\'endez et al. \cite{men08}). In these simulations the emphasis is on modelling the PNLF rather than explaining it and an alternative explanation for [O\,{\footnotesize III\normalsize}]-bright PNe is still required. Ciardullo et al. (\cite{cia05}) proposed that blue stragglers could evolve into [O\,{\footnotesize III\normalsize}]-bright PNe provided they are formed via a close binary interaction. Soker (\cite{sok06}) and Frankowski \& Soker (\cite{fra09}) suggested symbiotic stars could fulfill the role of [O\,{\footnotesize III\normalsize}]-bright PNe, but this would controversially require the majority of their nebulae to be ejected by the white dwarf (see Corradi \cite{cor03}). Both of these scenarios require alternative binary evolution channels for PN formation and as such they are not out of place amongst the growing evidence for binarity in PNe (De Marco \cite{dem09}). The frequency of PNe with close binary central stars that went through a common-envelope interaction is fairly high at $17\pm5$\% (Miszalski et al. \cite{mis09}) and symbiotic stars could potentially be rebranded as wide interacting binary central stars if we insist upon the nebula origin as already stated. Out of the two scenarios symbiotic stars are more readily accessible despite there being a severe paucity of Magellanic symbiotics (Belczy\'{n}ski et al. \cite{bel00}).

Magellanic PNe are well positioned to further advance our understanding of the PNLF. Large catalogues of [O\,{\footnotesize III\normalsize}] fluxes are available (Jacoby \& De Marco \cite{jac02}, Reid \& Parker \cite{rei10}) and the MCs are close enough to allow their PNe to be spatially resolved and studied in detail (e.g. Shaw et al. \cite{sha06}). This is a critical advantage over more distant [O\,{\footnotesize III\normalsize}] selected populations which are frequently assumed to contain only PNe. There are however a large variety of potential mimics that can contaminate the PNLF (see Frew \& Parker \cite{fre10} for a review) and symbiotic stars have been identified in the Local Group (Gon{\c c}alves et al. \cite{gon08}, Kniazev et al. \cite{kni09}). Ciardullo (\cite{cia10}) found a large scatter in the emission line ratios of objects in the top magnitude of the M 33 and LMC H$\alpha$ PNLFs which supports the case that more than one type of object can occupy the bright end. The deep near-IR photometry provided by VMC is sensitive to the dust associated with many mimics including compact HII regions and symbiotic stars. 

The synoptic nature of the VMC survey will detect variability due to Mira pulsations of the most obscured symbiotic stars that may otherwise be misclassified as PNe. These pulsations may not be visible at optical wavelengths (Miko{\l}ajewska et al. \cite{mik99}) providing a unique opportunity to increase the number of Magellanic symbiotic stars. We might also be sensitive to variability in the brightest central stars that could potentially be evidence for binarity if the variability is later found to be periodic (Miszalski et al. \cite{mis09}), however only a very small fraction of the population might be expected to fall within our magnitude limits (Villaver, Stanghellini \& Shaw \cite{vil07}).

\begin{figure}
\resizebox{\hsize}{!}{\includegraphics{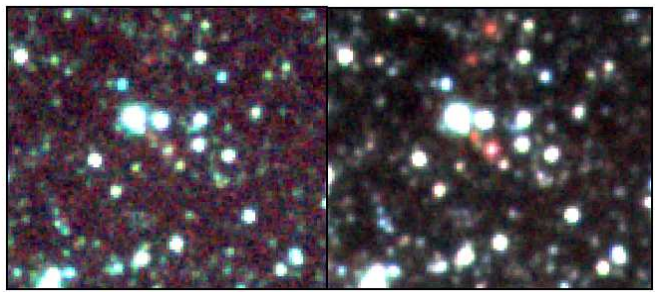}}
\caption{Colour-composite images of the LMC PN MG 60 made from single (left) and stacked (right) exposures of $Y$ (blue), $J$ (green) and $K_\mathrm{s}$ (red). The effective exposure times are $200$ sec for all single exposures, and $2400$ sec ($Y$ and $J$) and $6000$ sec ($K_\mathrm{s}$) for the stacked exposures. The images are $40\times35$ arcsec$^2$ with North up and East to the left.}
\label{fig:stacked}
\end{figure}

Within the first six VMC tiles in the LMC, a combination of optical imaging, 
OGLE light-curves, the VMC near-IR data, and SAGE mid-IR observations, 
reveals that only $\sim$50\% of the $98$ objects catalogued previously
as PNe appear to be genuine. These are characterised by the
colours $0.4\la J-K_\mathrm{s}\la2$ and $0.0\la Y-J\la0.5$. It is
encouraging though that almost all the genuine PNe are detected in all
wavebands and they appear especially bright in $K_\mathrm{s}$. The
strongest emission lines in the near-IR for PNe include He\,{\footnotesize I\normalsize}
$1.083\,\mu$m in $Y$, Pa $\beta$ in $J$, while $K_\mathrm{s}$ contains
Br $\gamma$, multiple He\,{\footnotesize I\normalsize} and molecular H$_2$ lines (Hora et
al. \cite{hor99}). Figure \ref{fig:stacked} shows the impact of
stacking the individual exposures to detect PNe, many of which are only
visible in the stacked fames. The non-PNe identified in the sample are mainly
misclassified field stars, compact H\,{\footnotesize II\normalsize} regions or long-period
variables. Our small fraction of genuine PNe is real and not just
accounted for by non-detections of faint nebulae in the near-IR. It is
also preliminary given the statistics are dominated heavily by the 30
Dor 6\_6 tile whose extensive H\,{\footnotesize II\normalsize} emission nebulosity complicates the
task of identifying PNe.

As the VMC survey becomes more complete, near-IR luminosity functions will be constructed and compared with their [O\,{\footnotesize III\normalsize}] (Jacoby \& De Marco \cite{jac02}; Reid \& Parker \cite{rei10}) and mid-IR (Hora et al. \cite{hor08}) counterparts. Improved diagnostic capabilities with the sensitivity of the deep $K_\mathrm{s}$ stacks and SAGE mid-IR photometry will also be applied to discover new PNe inside and especially outside the central $5\times5$ degrees covered by Reid \& Parker to create a more complete census of Magellanic PNe with accurate coordinates. 

\subsection{The stellar cluster KMHK 1577}

Stellar clusters are among the primary targets of the VMC survey. The
detection of known stellar clusters will be examined, and new clusters
will be searched for using a method similar to that already adopted in
the study of the MW (Ivanov et al. \cite{iva02}, Borissova et
al. \cite{bor03}). The analysis of stellar clusters will be centred on
the study of CMDs to estimate their ages and metallicities as well as
on the comparison with results obtained from optical surveys: MCPS
(Zaritsky et al. \cite{zar04}), OGLE (Pietrzynski et al. \cite{pie98},
\cite{pie99}), those by de Grijs \& Anders (\cite{deg06}) and other
dedicated studies. Ages, masses and metallicities of stellar clusters
will allows us to discuss the SFH of the MCs (e.g. synchronised bursts
at $\sim0.2$ and $2$ Gyr between the LMC and the SMC) and radial
abundance gradients. The first step of this work has been the
compilation of a catalogue of known stellar clusters located in the
Gaia SEP and 30 Dor fields from the list of Bica et
al. (\cite{bic08}).

\begin{figure*}
\resizebox{18cm}{5cm}{\includegraphics{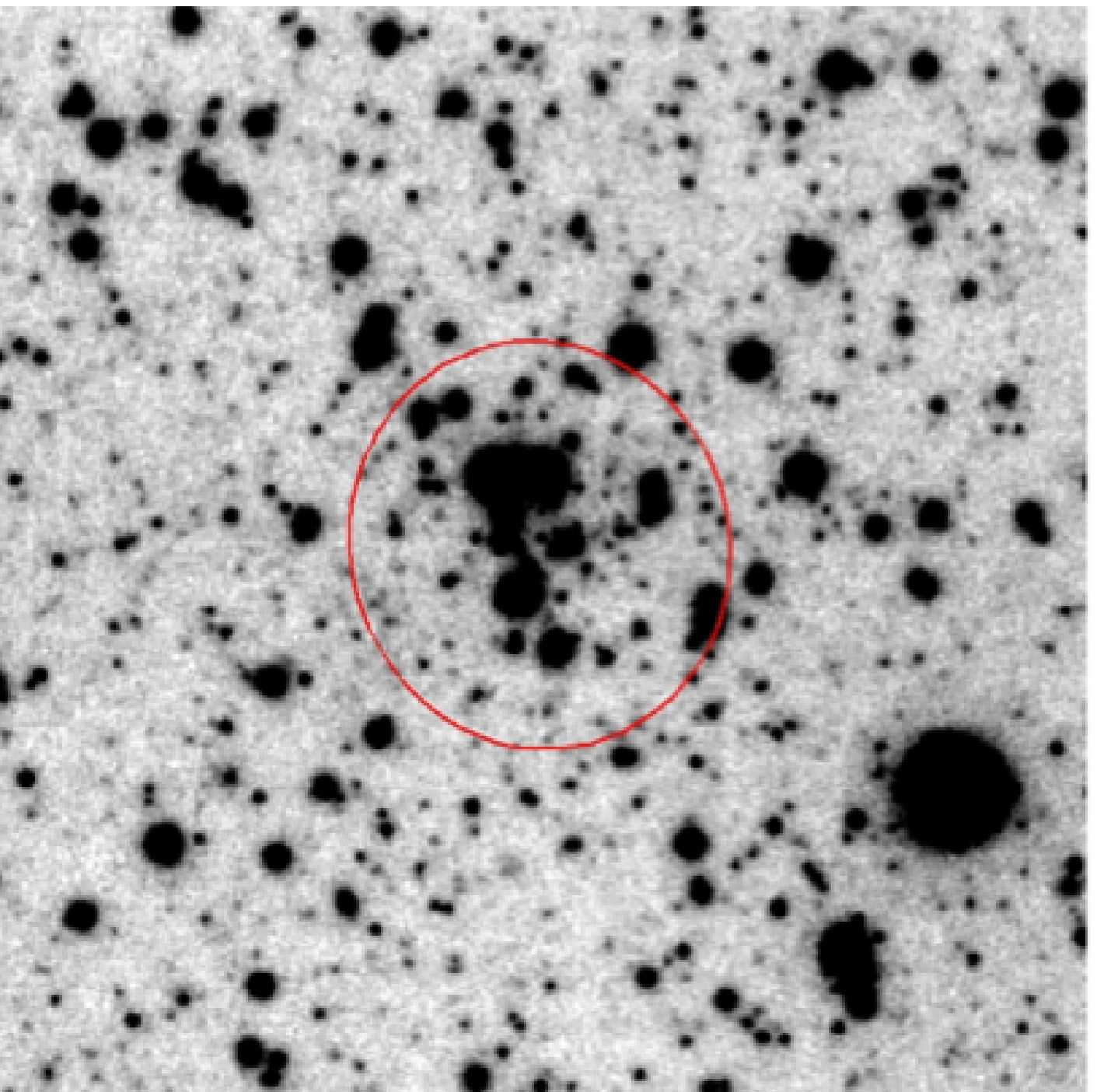}
\includegraphics{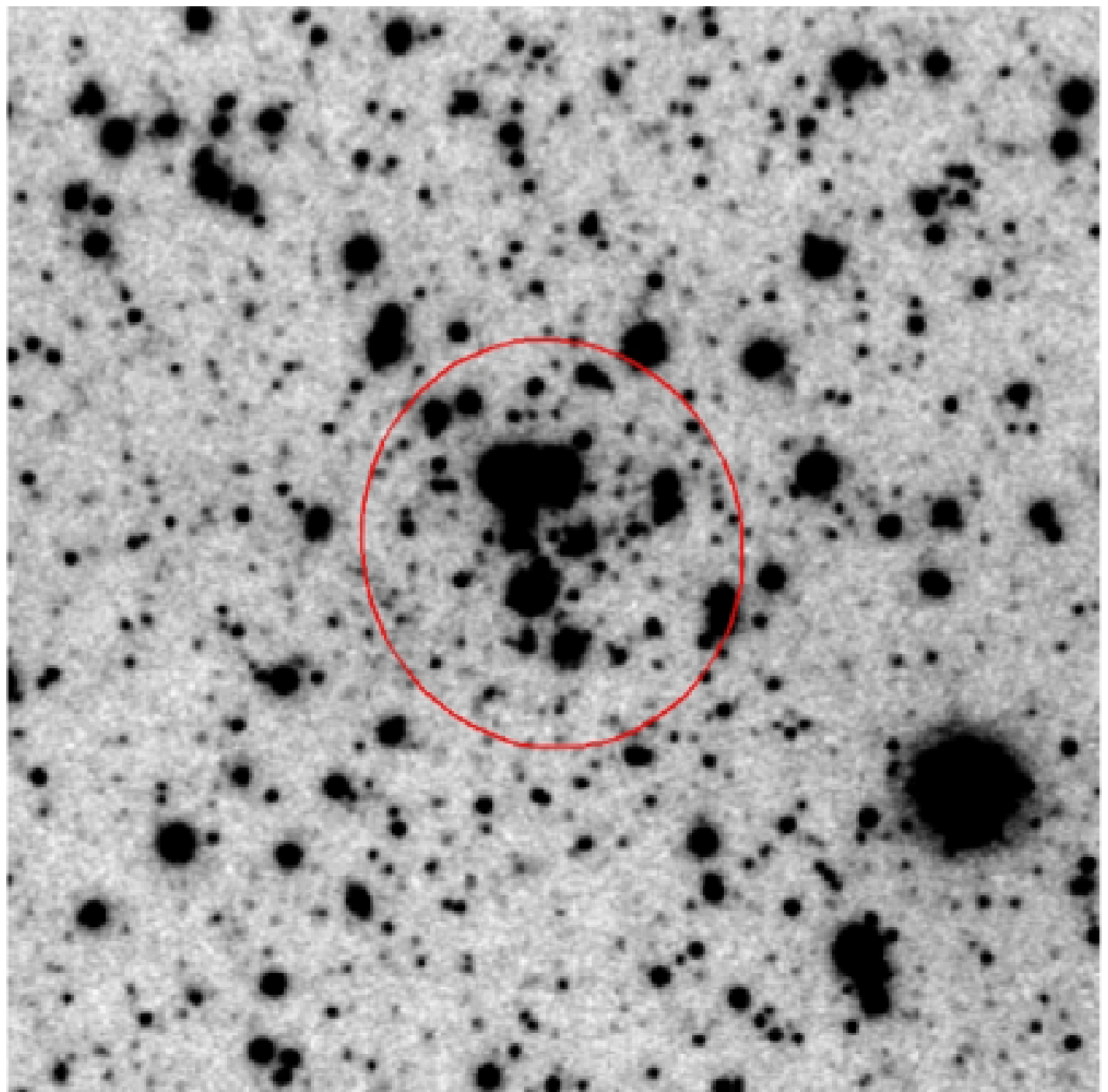}
\includegraphics{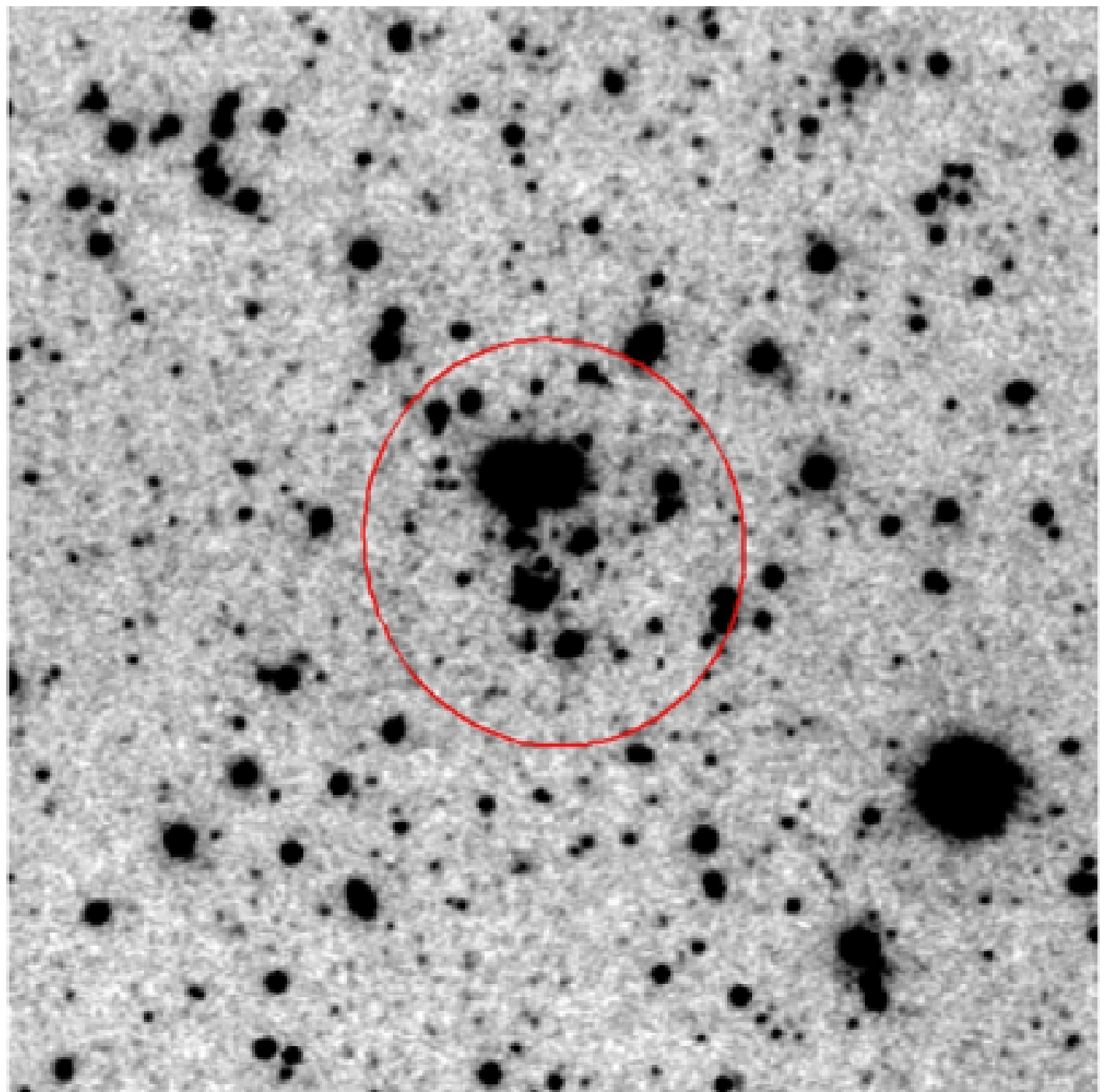}}
\caption{KMHK 1577 observed by VISTA in the $Y$, $J$ and $K_\mathrm{s}$ bands as part of the VMC survey: these images are about $2^\prime \times 2^\prime$ in size and refer to $1200$, $1400$ and $\sim 4000$ sec of integration, respectively. The ellipse shows the size of the cluster as indicated by Bica et al. (\cite{bic08}).} 
\label{fig1af}
\end{figure*}

In this paper we show preliminary results obtained from the study of
one cluster in the $8\_8$ field: KHMK 1577. This cluster appears in
the literature in different catalogues, including that from Bica et
al. (\cite{bic08}), but it is poorly studied and its properties (i.e. age
and metallicity) are unknown. It was chosen because of its favourable
location at the centre of a VMC paw-print which maximises the
availability of data for its stellar members. The VMC observations,
available in this field, are listed in Tabs. \ref{otab} and
\ref{ktab}. The CMD of the cluster and its immediate surroundings were
examined using photometry from deep VSA stacks. The three panels in
Fig. \ref{fig1af} show images of the cluster region in the $Y$, $J$
and $K_\mathrm{s}$ bands where the elliptical region occupied by KMHK
1577 is indicated.

The sources in the cluster region were divided into two concentric
groups: an inner group comprising all sources within an ellipse with a
major axis of $0.75^\prime$, a minor axis of $0.70^\prime$ and a
position angle of the major axis of $160^\circ$ (Bica et
al. \cite{bic08}), and an outer group of sources within a
circular area ($0.63^\prime$ in radius) equal in size to twice the
area of the inner ellipse. The stars within the elliptical region are
likely members of the cluster, although confirmation from radial
velocities would be necessary, while stars in the outer region are
unlikely associated with the cluster. By extracting the VMC data from
the VSA we find $69$ sources in the inner group and $98$ sources in
the outer group. If only sources detected in all three wave bands are
considered then there are $41$ and $55$ sources in the inner and outer
group, respectively. The inner group has a higher source
density. Using the VSA source classification flag there are $21$ stars
and $20$ galaxies in the inner group and $34$ stars and $21$ galaxies
in the outer group. Extended objects at faint magnitudes, however, are
most probably too faint for the classification criterion into stars
and galaxies to work. These objects may be elongated either because of
a low signal-to-noise or because they are blends. Their nature
will be clearer once all VMC epochs have been obtained and
stacked. Only one very bright object is classified as extended and
this may be the result of blending in the cluster centre. Objects with
red colours and low luminosities are likely extended objects
(galaxies, PNe and YSOs), none are present inside the ellipse.
Sources in the outer region, that are homogeneously distributed across
the area, are either giant stars of the LMC field population or
extended objects.

\begin{figure}
\resizebox{\hsize}{!}{\includegraphics{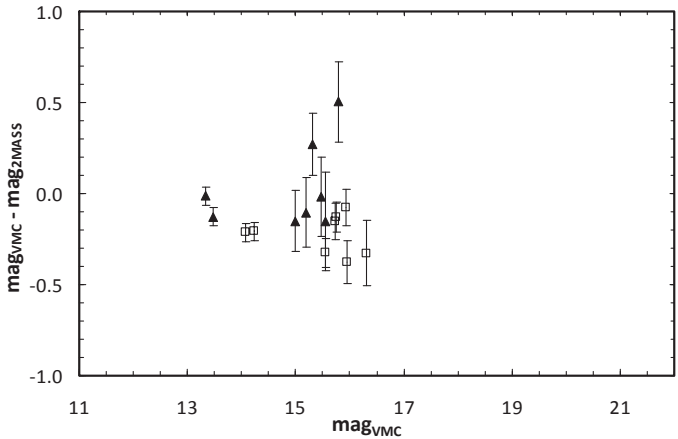}}
\caption{Magnitude differences between 2MASS and VMC detections in the $J$ (empty squares) and $K_\mathrm{s}$ (filled triangles) bands.}
\label{fig6ab}
\end{figure}

A comparison with 2MASS sources detected within the elliptical region
defining the cluster shows that at bright magnitudes
($K_\mathrm{s}\sim 13$) there is a very good agreement between the
2MASS and VMC magnitudes, with increasingly discrepant values when
approaching the 2MASS sensitivity limit. For the sources in common
between VMC and 2MASS (Fig. \ref{fig6ab}) the magnitude difference in
$J$ and $K_\mathrm{s}$ varies between being practically negligible to
values of $0.4-0.5$ mag. The difference in the $J-K_\mathrm{s}$ colour
is up to $0.8$ mag.

\begin{figure}
\resizebox{\hsize}{!} {\includegraphics{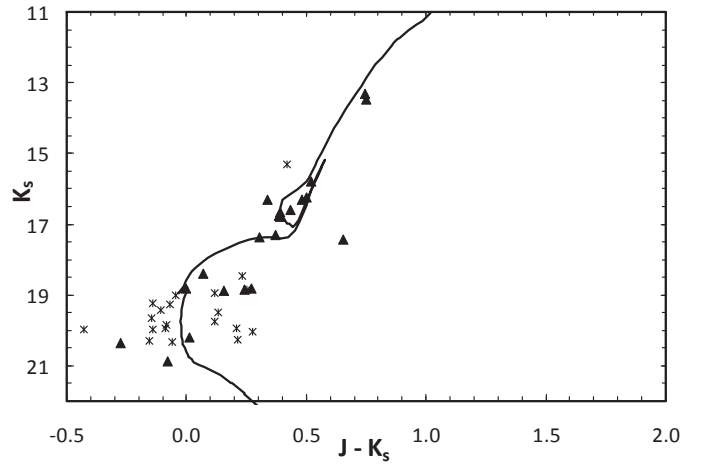}}
\caption{CMD ($J-K_\mathrm{s}$, $K_\mathrm{s}$) of the inner region around the stellar cluster KMHK 1577. Points represent stellar (filled triangles) and non-stellar (asterisks) objects. The line is the best fit isochrones from Marigo et al. (\cite{mar08}) with an age of $0.63$ Gyr and a metallicity of $Z=0.003$.}
\label{fig4af}
\end{figure}

The CMD shown in Fig. \ref{fig4af} indicates that the present sensitivity limit of VMC is at $K_\mathrm{s}\sim 21$ mag; this value agrees with expectations. Figure \ref{fig4af} shows that the entire RGB is well detected as well as the sub-giant branch and stars at the MS turn-off point. Photometric uncertainties vary with brightness and in the $K_\mathrm{s}$ band sources with a magnitude of $19-20$ have uncertainties of $ 0.2-0.4$ mag, sources with $16<K_\mathrm{s}<17$ have uncertainties of $0.01-0.04$ mag and the brightest sources at $K_\mathrm{s}\sim 13$ mag have an uncertainty of $\sim 0.003$ mag. $Y$ and $J$ bands sources have smaller photometric uncertainties than sources in $K_\mathrm{s}$. 

Using theoretical isochrones by Marigo et al. (\cite{mar08}) an age for the cluster of  $0.63 \pm 0.10$ Gyr and a metallicity of Z$=0.003\pm0.001$ were estimated. These values were derived from the average of the best fit isochrones found in three different CMDs: ($J-K_\mathrm{s}$, $K_\mathrm{s}$), ($Y-K_\mathrm{s}$, $K_\mathrm{s}$) and ($Y-J$, $J$). In each of these the best fit age was the same but the metallicity varied within the given dispersion. A similar procedure applied to the outer region gives an older ($1.4$ Gyr) and more metal poor (Z$=0.0004$) population. The absorption due to both foreground and interstellar extinction was estimated from $A_V=0.55$ mag (Zaritsky et al. \cite{zar04}) and the $A_\lambda / A_V$ ratios for an `average' LMC according to Gordon et al. (\cite{gor03}). This results in $A_Y=0.22$, $A_J=0.14$ and $A_{K_\mathrm{s}}=0.02$ mag where the $Y$ band value is an extrapolation from the nearest bands. The age of the cluster is typical of young LMC clusters while the low metallicity agrees with the cluster being located in a region quite far from the LMC centre.

%The field population in a control field centred $2^\prime$ South-East of KMHK 1577 at coordinates ($90.018^\circ$, $-66.793^\circ$) was examined. This field has the same size as for cluster KMHK 1577 and is essentially devoid of stars brighter than the sub-giant branch. 
%A best-fit isochrone indicates a population $\sim1.4$ Gyr old with a metallicity $Z=0.003$. While the metallicity is consistent with that of the KMHK 1577 cluster the age of the field population is older. 

\begin{figure}
\resizebox{\hsize}{!}{\includegraphics{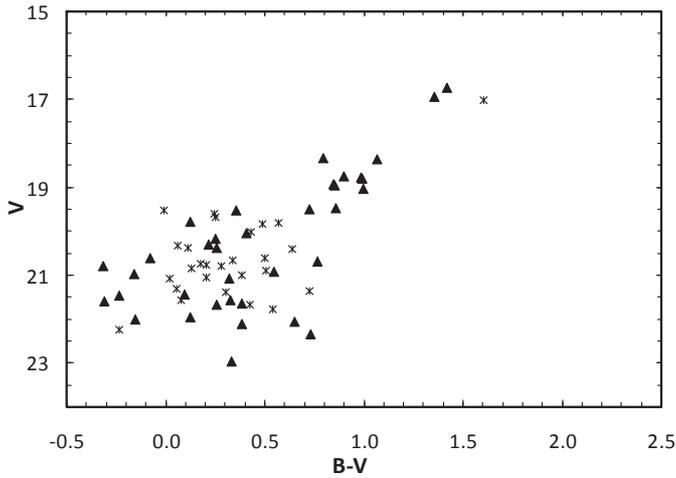}}
\caption{CMD ($V$, $B-V$) obtained from the MCPS data (Zaritsky et al. \cite{zar04}). Points, for VMC counterparts, are as in Fig. \ref{fig4af}.} 
\label{fig5f}
\end{figure}

Figure \ref{fig5f} shows a CMD ($B-V$, $B$) for the cluster KHMK 1577 from the MCPS data (Zaritsky et al. \cite{zar04}). With reference to the two groups analysed here we find $63$ sources in the inner group and $70$ sources in the outer group. Although it appears that MCPS has more stars in the central region of the cluster and allows for a better identification of the density enhancement, the VMC data delineate better features in the CMD (compare  Fig. \ref{fig4af} with Fig. \ref{fig5f}). Extinction and crowding effects in the optical data are the most likely explanation for the difference between the two diagrams. 

The VMC data harbour a great potential for investigating the evolutionary properties of stellar clusters. The internal spatial distribution of stars for each individual stellar cluster would provide information for dynamical studies.

\section{Conclusions}
\label{conclusions}

The VMC survey is a homogeneous and uniform $YJK_\mathrm{s}$ survey of
$\sim$184 deg$^2$ across the Magellanic system (Fig. \ref{tiles}). It
is an ESO Public Survey that started observations in
November $2009$ and will run for approximately five
years. The VMC survey parameters are described in
Tab. \ref{vmcparam}. Images and catalogues will be delivered to the
astronomical community at regular intervals with the first release
expected in 2011. The VMC data will provide, among other things, a
detailed history of star formation across the Magellanic system and a
measurement of its 3D geometry.

This paper presents the VMC survey strategy and first results aimed at
assessing their scientific quality. These show the potential of the
survey in addressing its main science goals and validates the expected
sensitivity of the VMC data. Colour-magnitude and colour-colour
diagrams show a wealth of substructures and a clear separation from
Galactic foreground stars. These diagrams will form the core of the
SFH analysis. To illustrate some of the scientific applications of the
VMC survey, Cepheids and RR Lyrae stars are shown to display clear
near-IR light-curves, PNe are detected thanks to the deep
$K_\mathrm{s}$ band images dominated by Br$\gamma$ emission, and
stellar cluster parameters are derived from best fit isochrones to the
VMC colour-magnitude diagrams.

The VMC survey will be of immense value to the astronomical community
because the data will represent the only counterpart for existing
optical surveys at a similar sensitivity (e.g. MCPS) and for the large
number of unclassified objects observed by the {\em Spitzer Space
Telescope} in the mid-IR (Blum et al. \cite{blu06}). Note that the
near-IR 2MASS survey has observed only about 6\% of the stars that the
VMC survey is expected to detect. The VMC data cover the bulk
of the Magellanic system, as opposed to the tiny regions sampled by
the Hubble Space Telescope, and the limited area covered by most of
the other dedicated, ground-based observations at the same sensitivity.

Among the other VISTA Public Surveys, the VISTA Hemisphere Survey
(VHS) will also contribute to the investigation of the Magellanic
system. The VHS is $\sim$3 mag shallower than the VMC survey but it
covers an extended area around the Magellanic system to complement
the VMC data.

\begin{acknowledgements}
BM thanks George Jacoby for sharing his database of SMC PNe. MG and MATG acknowledge financial support from the Belgian Federal Science Policy (project MO/33/026). We thank the UK team responsible for the realisation of VISTA, and the ESO team who have been operating and maintaining this new facility. The UK's VISTA Data Flow System comprising the VISTA pipeline at CASU and the VISTA Science Archive at WFAU has been crucial in providing us with calibrated data products for this paper, and is supported by STFC. This research has made use of Aladin, EROS-2 data which were kindly provided by the EROS collaboration and of the Southern H-Alpha Sky Survey Atlas (SHASSA), which is supported by the National Science Foundation. This work was partially supported by  PRIN-INAF 2007: `Resolved stellar populations in the near-, mid- and far-infrared (P.I. L. Girardi)' and PRIN-INAF 2008: `The ESO Magellanic Cloud Surveys: tracing the stellar populations and beyond' (P.I. M. Marconi).
\end{acknowledgements}

\begin{appendix}
\label{app}

\section{VMC tile centres}

\begin{figure}
\resizebox{\hsize}{!}{\includegraphics[angle=270]{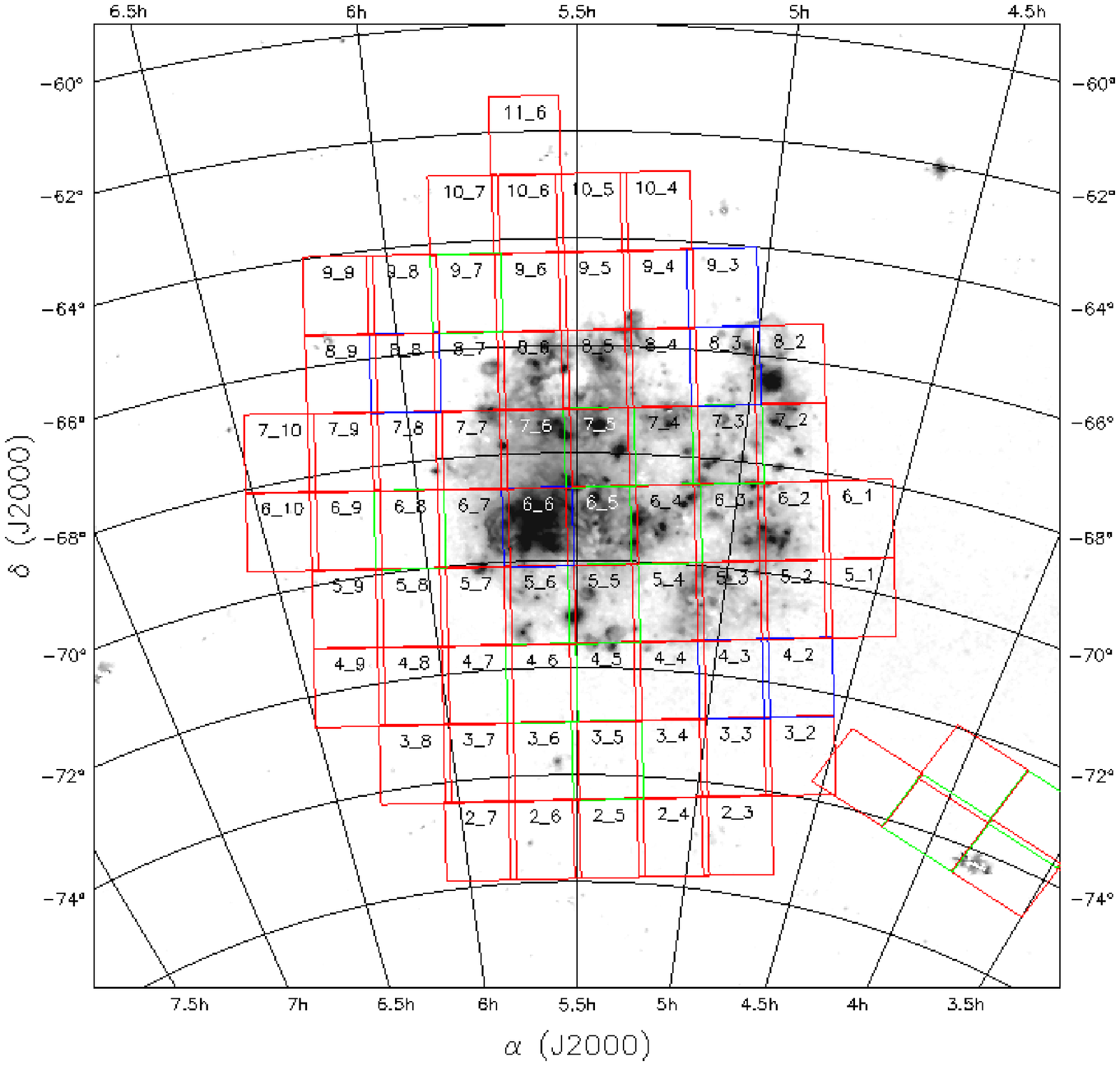}}
\caption{LMC area tiled for VMC observations. The underlying image shows the H$\alpha$ distribution (Gaustad et al. \cite{gau01}). VISTA tiles are colour-coded as in Fig. \ref{tiles}.}
\label{tileslmc}
\end{figure}

\begin{figure}
\resizebox{\hsize}{!}{\includegraphics[angle=270]{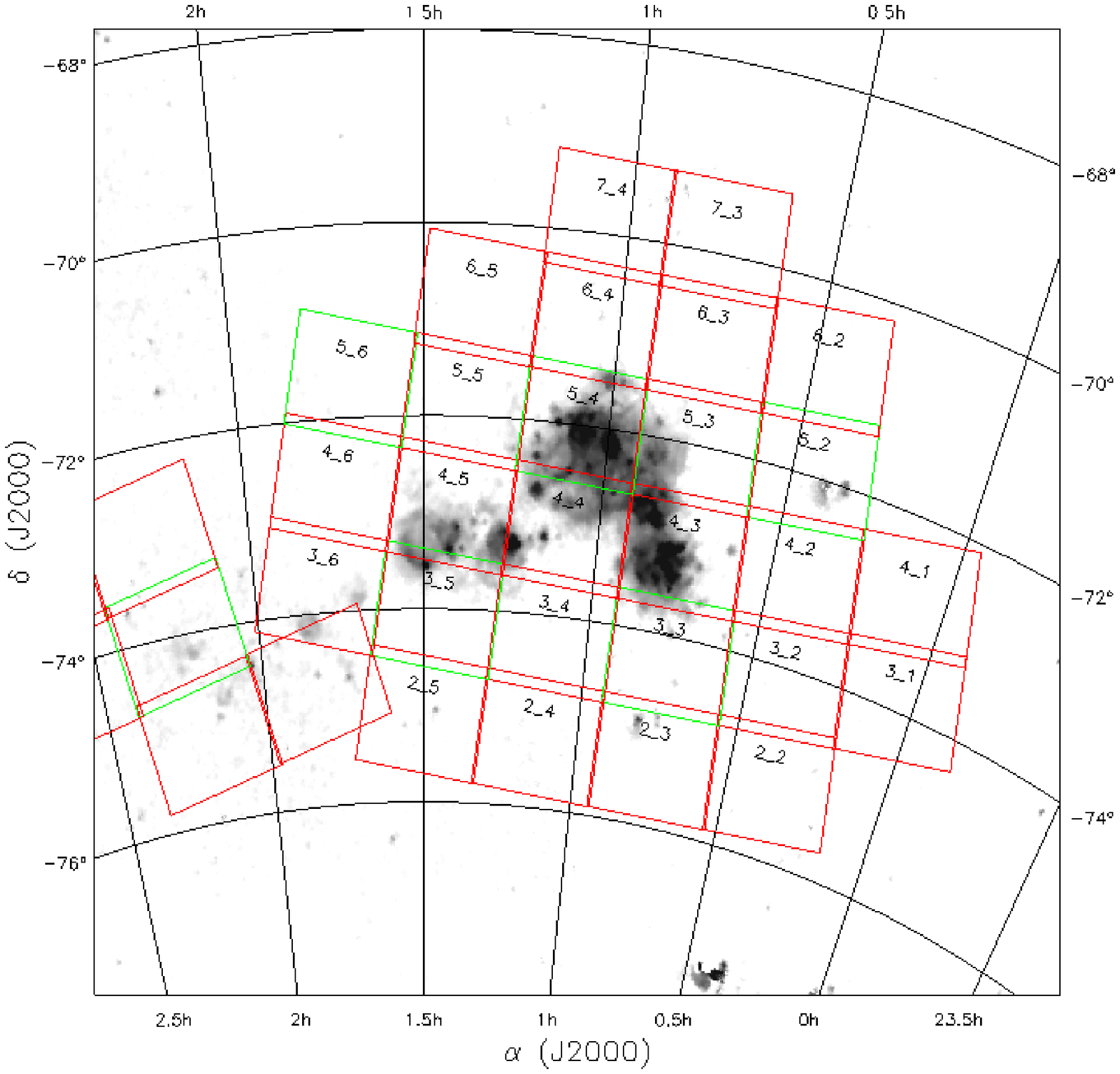}}
\caption{SMC area tiled for VMC observations. The underlying image shows the H$\alpha$ distribution (Gaustad et al. \cite{gau01}). VISTA tiles are colour-coded as in Fig. \ref{tiles}.}
\label{tilessmc}
\end{figure}

\begin{figure}
\resizebox{\hsize}{!}{\includegraphics[angle=270]{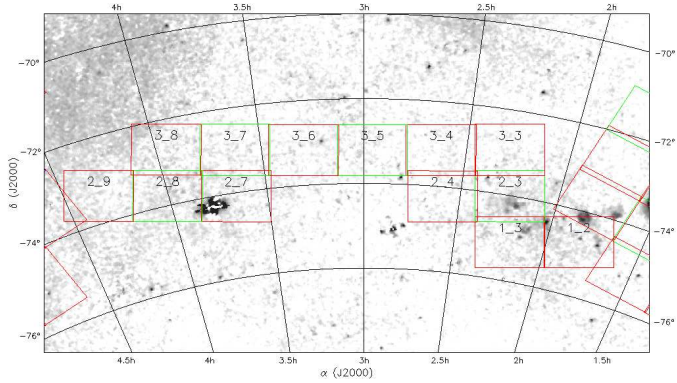}}
\caption{Bridge area tiled for VMC observations. The underlying image shows the H$\alpha$ distribution (McClure-Griffiths et al. \cite{mcc09}). VISTA tiles are colour-coded as in Fig. \ref{tiles}.}
\label{tilesbridge}
\end{figure}

Tables \ref{lmccentres1} and \ref{lmccentres2} show the centres of VMC tiles covering the LMC, while Table \ref{smccentres} shows the centres of tiles covering the SMC and Table \ref{bridgecentres} shows the centres of tiles covering the Bridge. The tile identification is formed of two numbers. The first number indicates the row and the second number the column that correspond to the location of a given tile. Row numbers increase from bottom to top while column numbers increase from right to left. Refer to Figures \ref{tileslmc}, \ref{tilessmc} and \ref{tilesbridge} for the location of tiles across the LMC, SMC and Bridge components of the Magellanic system, respectively. For the Stream the central coordinates of two distinct tiles are indicated in Table \ref{streamcentres}. Note that since tiles were generated from almost rectangular grids, some outer tiles were removed leading to the tile numbers shown here.

\begin{table}
\caption{LMC tile centres}
\label{lmccentres1}
\[
\begin{array}{cccl}
\hline
\noalign{\smallskip}
\mathrm{Tile} & \alpha & \delta & \mathrm{Comments}\\
\hline
\noalign{\smallskip}
2\_3  &    04:48:04.752     &       -74:54:11.880 &\\
2\_4  &    05:04:42.696     &       -75:04:45.120 &\\
2\_5  &    05:21:38.664     &       -75:10:50.160 &\\
2\_6  &    05:38:43.056     &       -75:12:21.240 &\\
2\_7  &    05:55:45.720     &       -75:09:17.280 &\\
3\_2  &    04:37:05.256     &       -73:14:30.120 &\\
3\_3  &    04:51:59.640     &       -73:28:09.120 &\\
3\_4  &    05:07:14.472     &       -73:37:49.800 &\\
3\_5  &    05:22:43.056     &       -73:43:25.320 & \mathrm{started} \\
3\_6  &    05:38:18.096     &       -73:44:51.000 &\\
3\_7  &    05:53:51.912     &       -73:42:05.760 &\\
3\_8  &    06:09:16.920     &       -73:35:12.120 &\\
4\_2  &    04:41:30.768     &       -71:49:16.320 & \mathrm{started} \\
4\_3  &    04:55:19.512     &       -72:01:53.400 & \mathrm{started} \\
4\_4  &    05:09:24.288     &       -72:10:49.800 &\\
4\_5  &    05:23:39.816     &       -72:15:59.760 &\\
4\_6  &    05:38:00.408     &       -72:17:20.040 & \mathrm{started} \\
4\_7  &    05:52:20.064     &       -72:14:49.920 &\\
4\_8  &    06:06:32.952     &       -72:08:31.200 &\\
4\_9  &    06:20:33.408     &       -71:58:27.120 &\\
5\_1  &    04:32:43.848     &       -70:08:40.200 &\\
5\_2  &    04:45:19.440     &       -70:23:43.800 &\\
5\_3  &    04:58:11.664     &       -70:35:27.960 &\\
5\_4  &    05:11:16.704     &       -70:43:46.200 &\\
5\_5  &    05:24:30.336     &       -70:48:34.200 & \mathrm{started} \\
5\_6  &    05:37:47.952     &       -70:49:49.440 &\\
5\_7  &    05:51:04.872     &       -70:47:31.200 &\\
5\_8  &    06:04:16.416     &       -70:41:40.560 &\\
5\_9  &    06:17:18.096     &       -70:32:20.760 &\\
6\_1  &    04:36:49.488     &       -68:43:50.880 &\\
6\_2  &    04:48:39.072     &       -68:57:56.520 &\\
6\_3  &    05:00:42.216     &       -69:08:54.240 &\\
6\_4  &    05:12:55.800     &       -69:16:39.360 & \mathrm{started} \\
6\_5  &    05:25:16.271     &       -69:21:08.280 &\\
6\_6  &    05:37:40.008     &       -69:22:18.120 & \mathrm{30\,Dor\,-\,completed} \\
6\_7  &    05:50:03.168     &       -69:20:09.240 &\\
6\_8  &    06:02:21.984     &       -69:14:42.360 & \mathrm{in\, queue} \\
6\_9  &    06:14:32.832     &       -69:05:59.640 &\\
6\_10 &   06:26:32.280     &       -68:54:05.760 &\\
7\_2  &    04:51:34.992     &       -67:31:57.000 &\\
7\_3  &    05:02:55.200     &       -67:42:14.760 & \mathrm{in\, queue} \\
7\_4  &    05:14:23.976     &       -67:49:30.720 &\\
7\_5  &    05:25:58.440     &       -67:53:42.000 & \mathrm{in\, queue} \\
7\_6  &    05:37:35.544     &       -67:54:47.160 &\\
7\_7  &    05:49:12.192     &       -67:52:45.480 &\\
7\_8  &    06:00:45.240     &       -67:47:38.040 &\\
7\_9  &    06:12:11.736     &       -67:39:26.640 &\\
7\_10 &  06:23:28.800     &       -67:28:14.880 &\\
\noalign{\smallskip}
\hline
\end{array}
\]
\end{table}

\begin{table}
\caption{LMC tile centres (continue)}
\label{lmccentres2}
\[
\begin{array}{cccl}
\hline
\noalign{\smallskip}
\mathrm{Tile} & \alpha & \delta & \mathrm{Comments} \\
\hline
\noalign{\smallskip}
8\_2  &    04:54:11.568     &       -66:05:47.760 &\\
8\_3  &    05:04:53.952     &       -66:15:29.880 & \mathrm{started} \\
8\_4  &    05:15:43.464     &       -66:22:19.920 &\\
8\_5  &    05:26:37.704     &       -66:26:15.720 &\\
8\_6  &    05:37:34.104     &       -66:27:15.840 &\\
8\_7  &    05:48:30.120     &       -66:25:19.920 &\\
8\_8  &    05:59:23.136     &       -66:20:28.680 & \mathrm{Gaia\,-\,completed} \\
8\_9  &    06:10:10.632     &       -66:12:43.560 &\\
9\_3  &    05:06:40.632     &       -64:48:40.320 & \mathrm{started} \\
9\_4  &    05:16:55.464     &       -64:55:07.680 &\\
9\_5  &    05:27:14.256     &       -64:58:49.440 &\\
9\_6  &   05:37:34.872     &       -64:59:44.520 &\\
9\_7  &    05:47:55.128     &       -64:57:52.920 & \mathrm{in\,queue}\\
9\_8  &    05:58:12.816     &       -64:53:15.000 &\\
9\_9  &    06:08:25.848     &       -64:45:52.560 &\\
10\_4 &  05:18:01.536     &       -63:27:54.000 &\\     
10\_5 &   05:27:48.912     &       -63:31:22.800 &\\
10\_6 &   05:37:37.800     &       -63:32:13.200 &\\
10\_7 &   05:47:26.352     &       -63:30:24.840 &\\
11\_6 &   05:37:42.432     &       -62:04:41.520 &\\
\noalign{\smallskip}
\hline
\end{array}
\]
\end{table}

\begin{table}
\caption{SMC tile centres}
\label{smccentres}
\[
\begin{array}{cccl}
\hline
\noalign{\smallskip}
\mathrm{Tile} & \alpha & \delta & \mathrm{Comments} \\
 \hline
\noalign{\smallskip}
2\_2	& 00:21:43.920  &  -75:12:04.320 & \\
2\_3  & 00:44:35.904 & -75:18:13.320 & \\
2\_4  & 01:07:33.864  & -75:15:59.760 & \\
2\_5  & 01:30:12.624  & -75:05:27.600 & \\
3\_1  & 00:02:39.912  & -73:53:31.920 & \\
3\_2  & 00:23:35.544  & -74:06:57.240 & \\ 
3\_3  & 00:44:55.896  & -74:12:42.120 & \mathrm{started} \\
3\_4  & 01:06:21.120  & -74:10:38.640 & \\
3\_5  & 01:27:30.816  & -74:00:49.320 & \mathrm{started} \\
3\_6  & 01:48:06.120  & -73:43:28.200 & \\
4\_1  & 00:05:33.864  & -72:49:12.000 & \\
4\_2  & 00:25:14.088  & -73:01:47.640 & \\
4\_3  & 00:45:14.688  & -73:07:11.280 & \\
4\_4  & 01:05:19.272  & -73:05:15.360 & \mathrm{NGC\,419} \\
4\_5  & 01:25:11.088  & -72:56:02.760 & \\
4\_6  & 01:44:34.512  & -72:39:44.640 & \\
5\_2  & 00:26:41.688  & -71:56:35.880 & \mathrm{started} \\
5\_3  & 00:45:32.232  & -72:01:40.080 & \\
5\_4  & 01:04:26.112  & -71:59:51.000 & \mathrm{NGC\,411\,-\,started} \\
5\_5  & 01:23:09.336  & -71:51:09.720 & \\
5\_6  & 01:41:28.800  & -71:35:47.040 & \mathrm{in\,queue} \\
6\_2  & 00:28:00.192  & -70:51:21.960 & \\
6\_3  & 00:45:48.792  & -70:56:09.240 & \\
6\_4  & 01:03:40.152  & -70:54:25.200 & \\
6\_5  & 01:21:22.560  & -70:46:11.640 & \\
7\_3  & 00:46:04.728  & -69:50:38.040 & \\
7\_4  & 01:03:00.480  & -69:48:58.320 & \\
\noalign{\smallskip}
\hline
\end{array}
\]
\end{table}

\begin{table}
\caption{Bridge tile centres}
\label{bridgecentres}
\[
\begin{array}{cccl}
\hline
\noalign{\smallskip}
\mathrm{Tile} & \alpha & \delta & \mathrm{Comments} \\
\hline
\noalign{\smallskip}
1\_2  & 01:49:51.960  & -74:43:31.800 & \\
1\_3  & 02:11:35.232  & -75:05:04.560 & \\
2\_3  & 02:14:46.584  & -74:00:47.520 & \mathrm{started} \\
2\_4  & 02:35:28.440  & -74:13:18.840 & \\
2\_7  & 03:39:43.800  & -74:04:51.960 & \\
2\_8  & 04:00:21.072  & -73:46:37.560 & \mathrm{started} \\
2\_9  & 04:20:05.640  & -73:21:14.040 & \\
3\_3  & 02:17:35.496  & -72:56:22.200 & \\
3\_4  & 02:37:24.888  & -73:08:16.440 & \\
3\_5  & 02:57:33.288  & -73:12:52.200 & \mathrm{started} \\
3\_6  & 03:17:43.776  & -73:10:03.720 & \\
3\_7  & 03:37:39.240  & -72:59:54.600 & \mathrm{in\,queue} \\
3\_8  & 03:57:03.888  & -72:42:37.800 & \\
\noalign{\smallskip}
\hline
\end{array}
\]
\end{table}

\begin{table}
\caption{Stream tile centres}
\label{streamcentres}
\[
\begin{array}{cccl}
\hline
\noalign{\smallskip}
\mathrm{Tile} & \alpha & \delta & \mathrm{Comments}\\
\hline
\noalign{\smallskip}
1\_1 & 03:30:03.87 & -64:25:24.8 & \mathrm{started}\\
2\_1 & 00:11:59.30 & -64:39:31.8 & \mathrm{started}\\
\noalign{\smallskip}
\hline
\end{array}
\]
\end{table}

\end{appendix}

\end{document}